# Effectively integrating information content and structural relationship to improve the GO-based similarity measure between proteins

Bo Li[1], James Z. Wang[1], F. Alex Feltus[2], Jizhong Zhou[3], Feng Luo[1, 2, §]

[1]School of Computing, Clemson University, Clemson, SC, USA

[2]Department of Biochemistry and Genetics, Clemson University, Clemson, SC, USA

[3]Institute for Environmental Genomics, and Department of Botany and Microbiology, University of Oklahoma, Norman, OK, 73019, USA,

[§]Corresponding author

Email addresses:

    BL: bol@clemson.edu

    JZW: jzwang@cs.clemson.edu

    FAF: ffeltus@clemson.edu

    JZZ: jzhou@ou.edu

    FL: luofeng@clemson.edu

# Abstract

**Background**

The Gene Ontology (GO) provides a knowledge base to effectively describe proteins. However, measuring similarity between proteins based on GO remains a challenge.

**Results**

In this paper, we propose a new similarity measure, information coefficient similarity measure ($Sim_{IC}$), to effectively integrate both the information content (IC) of GO terms and the structural information of GO hierarchy to determine the similarity between proteins. Testing on yeast proteins, our results show that $Sim_{IC}$ efficiently addresses the shallow annotation issue in GO, thus improves the correlations between GO similarities of yeast proteins and their expression similarities as well as between GO similarities of yeast proteins and their sequence similarities. Furthermore, we demonstrate that the proposed $Sim_{IC}$ is superior in predicting yeast protein interactions. We predict 20484 yeast protein-protein interactions (PPIs) between 2462 proteins based on the high $Sim_{IC}$ values of biological process (BP) and cellular component (CC). Examining the 214 MIPS complexes in our predicted PPIs shows that all members of 159 MIPS complexes can be found in our PPI predictions, which is more than those (120/214) found in PPIs predicted by relative specificity similarity (RSS).

**Conclusions**

Integrating IC and structural information of GO hierarchy can improve the effectiveness of the semantic similarity measure of GO terms. The new $Sim_{IC}$ can effectively correct the effect of shallow annotation, and then provide an effective way to measure similarity between proteins based on Gene Ontology.

# Background

The Gene Ontology [1] is a database of controlled vocabulary terms that represent human knowledge about genes and gene products. GO terms are divided into three categories: biological process (BP), molecular function (MF), and cellular component (CC). The cellular component terms characterize the location of gene products in the cell. The molecular function terms represent the molecular level activities of proteins. The biological process terms describe a series of events accomplished by one or more proteins [1]. The internal GO structure is a hierarchical directed acyclic graph (DAG), in which the terms are vertices and the relationships form the edges. Each of the three categories forms a branch of the GO DAG. There are five types of semantic relationships in the GO DAG, and we focus our study on two of them, namely, "is_a", which implies that the child is a subclass of the parent, and "part_of", which indicates that the child is component of the parent. Currently, the GO is an important knowledge resource to describe proteins, and using GO annotations to measure the similarity between gene products has attracted a lot of attention [2-20]. In these approaches, the similarity between two proteins is calculated based on semantic similarities among GO terms that annotate them.

There are many challenges in the measurement of semantic similarity between GO terms and between proteins. First, the depth of empirical knowledge and the number of associated GO terms on each aspect of a given protein is highly variable. This creates a situation where the hierarchical level of annotated GO terms may not accurately represent the underlying concept that GO terms at the same level in the DAG may not be equally specific [7]. Second, annotating gene products with GO terms is often limited by bioinformatics techniques including the sensitivity of homology detection for functional mapping. Proteins may be annotated with high level GO terms which are close to the root of the DAG or with GO terms associated with a large number of genes thereby diluting functional information.

This phenomenon is called "shallow" annotation. Furthermore, a large number of gene products have not been annotated with any GO terms at all.

The similarity measures between GO terms are usually based on two factors: information content (IC) of GO terms and structural information of GO hierarchy. A GO term is less informative if it is frequently used to annotate gene products. That is, the information content of a GO term is inversely proportional to its probability of correctly annotating genes. The probability of a given term *t* is defined as the number of genes annotated with it over the total number of annotations. As a gene that is annotated with a GO term is also annotated with all ancestors of the GO term in the DAG, the total number of annotations is the frequency of DAG root. Namely, the probability of a term t is given by:

$$prob(t) = \frac{anno(t) + \sum_{d \in descendant(t)} anno(d)}{\sum_{c \in descendant(root)} anno(c)} \quad (1)$$

where anno(t) is the number of genes directly annotated with the term t. The hierarchical structure of GO makes the probability of GO terms non-decreasing from the descendants to the root of the DAG and the probability of the DAG root has a value of 1. The IC of a GO term is usually defined as negative logarithm of the term's probability [21]:

$$IC(t) = -\log(prob(t)) \quad (2)$$

Thus, the information content of DAG root is 0. As GO terms ascend the hierarchical tree of DAG from the leaf, the information content would not increase.

Commonly used structural information of GO hierarchy is the topological path between two GO terms, which is the number of steps taken to reach one term from another in the DAG. However, the same distance between two GO terms close to the root of the DAG (more general) and between two GO terms close to the leaf of the DAG (more specific) doesn't mean the similarities of these two GO term pairs are conceptually equal. Furthermore, the GO terms in the same level may also not be of the same specificity. Thus, the path between two GO terms is not the ideal similarity measure. Meanwhile, two structural concepts that are frequently used in GO similarity measures are the "lowest common ancestor" (LCA) and "most informative common ancestor" (MICA) [21]. Given two GO terms, the MICA is their common ancestor with the lowest probability in the DAG structure. In the example of gene ontology DAG shown in Figure 1, there are three common ancestors of terms H and I which are A, D and F. As term A definitely has higher probability than terms D and F have. The MICA of term H and I is one of terms D and F, which should have lower probability than term A. The LCA of two GO terms is their common ancestor term at the lowest level in the DAG structure. As shown in Figure 1, the LCA of terms H and I is F. Notice that there also exists a case where two terms have multiple ancestor terms at the same level. In this case, the term with lowest probability will be the final LCA. The relationship between two terms and their LCA or MICA has been widely used to designing the similarity measures.

Many similarity measures have been developed based on structural relationship, information content [13, 21-26] or both. Most of these similarity measures are originally designed for other ontologies, such as WordNet [27]. Unfortunately, most existing similarity measures have not effectively integrated information content and structural information. Some similarity measures (e.g. Resnik [21]) employ information content of the MICA of two concepts only. Although these measures can reflect the semantic difference among the DAG levels, they lose the structural information in the DAG and cannot reveal the semantic difference among term pairs with same MICA. As shown in Figure 1, both the (L, D) pair and the (L, K) pair have the same MICA, B. Thus, they will have the same Resnik similarity. However, conceptually, L should be more similar to D. Other similarity measures such as Lin [23], Jiang [22], and graph information content (GIC) of Pesquita *et al* [26] (see Methods section for detailed definitions) utilize both information content and structural information, but they actually can not distinguish semantic differences among different levels in DAG when both proteins are annotated by the same GO term(s), in which case the GO term similarity is just the protein similarity. For the example of Figure 1, if two proteins are uniquely annotated by term J and two proteins are uniquely annotated by term C, these two pairs of proteins will have same similarities in Lin, Jiang and GIC, which is misleading. On the other hand, the Resnik measure can give different values in this case. Similarity measures, like the Relative Specificity Similarity (RSS) of Wu et al. [28] and Wang et al. [13], consider the structure information of the GO DAG only. These similarities are also affected by shallow annotations because no information content is incorporated into them. For the example of Figure 1, if a pair of proteins (pair A) are only annotated by GO term J and another pair of proteins (pair B) are only annotated by GO term C, the Wang measure will give both pairs of proteins with similarity 1.0 and can not distinguish the semantic difference of GO terms J and C. The RSS is also insufficient to measure the similarity between protein pairs (see Results sections for more discussion).

Recently, Schlicker et al. [24] proposed a relevance similarity that counts both the structural relationship between two terms and their MICA and the probability of the MICA. The relevance similarity is defined as the product of Lin similarity and a coefficient, (1-prob(MICA)). The adjustment of Lin similarity with (1-prob(MICA)) can reduce the effect of shallow annotations. Unfortunately, the probability adjustment becomes insensitive when the probability value is close to 0 and 1. For example, if there are about ten thousand biological process annotations, a GO term with 100 annotations will have a probability of 0.01. On the other hand, a GO term with ten annotations will have a probability of 0.001. Then, the adjustments in relevance similarity are 0.99 and 0.999 respectively. These adjustments are not sensitive enough to detect their information difference, which is a ten-fold difference in annotation number.

In this paper, we propose a new method to calculate similarity between two GO terms by effectively integrating structural information with IC. Our $Sim_{IC}$ similarity measure is sensitive to IC differences among GO terms and efficiently addresses the shallow annotation problem. We have compared $Sim_{IC}$ with Lin, Jiang and GIC, relevance, Resnik and adjusted Resnik similarities by calculating the correlations between semantic similarities of yeast protein pairs and their gene expression similarities and correlations between semantics similarities of yeast protein pairs and sequence similarities. All comparisons are based on the GO annotations of yeast from Saccharomyces Genome Database (SGD) [29]. Furthermore, we show that $Sim_{IC}$ is superior in predicting yeast protein interactions.

## Results
**New semantic similarity measures by integrating IC and structure of GO DAG**

The relevance similarity proposed by Shlicker et al. [24] is not sensitive to the semantic difference among GO term close to the leaf of DAG (probability close to 0) or close to the root of the DAG (probability close to 1). In order to enforce more impact of information content on the similarity detection, we introduce a new information coefficient similarity ($Sim_{IC}$) measure to effectively integrate information content and structural relationship. The $Sim_{IC}$ is defined as:

$$Sim_{IC}(t_1, t_2) = \frac{2 \times \log(prob(MICA))}{\log(prob(t_1)) + \log(prob(t_2))} \times (1 - \frac{1}{1 + IC(MICA)}) \qquad (3)$$

Using the example in background section, the $Sim_{IC}$ adjustments for the GO term with 100 annotations and the GO term with ten annotations are 0.67 and 0.75 respectively. The adjustment difference is 0.08, which is much higher than the difference between probability adjustments used in relevance similarity, which is 0.009. The first part of $Sim_{IC}$ is simply the Lin similarity. We also can obtain other flavours of $Sim_{IC}$ by replacing the first part of $Sim_{IC}$ with Jiang or other measures. Unless explicitly stated otherwise, we will refer $Sim_{IC}$ as the one defined in Equation 3.

**Correction of the shallow annotation by SimIC**

Many proteins are annotated by general GO terms due to limited knowledge about these proteins and their homologs. As discussed in Background section, these shallow annotations will affect the accuracy of similarity measures, such as Lin, Jiang, Wang and GIC measures. Both the relevance similarity and $Sim_{IC}$ are able to correct the shallow annotation effect using information content. Comparing to the relevance similarity, the $Sim_{IC}$ will be provide more efficient correction for the GO terms close to root. Figure 2 shows the distributions of Lin, relevance and $Sim_{IC}$ of all pairs of annotated yeast proteins. As each similarity measure has different range of values, we take a normalization step by evenly partitioning values of each similarity measure into 21 bins. As shown in Figure 2, the $Sim_{IC}$ has a higher percentage of low values relative to Lin and relevance similarities. Moreover, the relevance similarity has similar percentage of the highest values (last bin) as the Lin similarity has, which means that the probability adjustment is limited to the highest values. On the other hand, the $Sim_{IC}$ is able to efficiently adjust the high similarity values caused by shallow annotations and lead to a very low percentage of highest values as shown in inner figure of Figure 2. For example, as shown in Figure 3A, yeast proteins YDR387C and YHL035C both annotated with term "transport", a level-three GO term with 1726 annotations in total associated with the term and its descendants. And as shown in Figure 3B, yeast proteins YNL108C and YJR051W both are annotated with "metabolic process", a level-two GO term with 7027 annotations in total associated with the term and its descendants. Both pairs of yeast proteins will have similarity values of 1.0 based on Lin, Wang, Jiang and GIC measures. The relevance similarity has a value of 0.8512 for YDR387C and YHL035C and a value of 0.5689 for YNL108C and YJR051W. Meanwhile, the $Sim_{IC}$ has a value of 0.6829 for YDR387C and YHL035C and a value of 0.4284 for YNL108C and YJR051W. Figure 4 compares $Sim_{IC}$ similarities with relevance similarities for yeast protein pairs with Lin's similarities greater than 0.9. While relevance similarities have values between 0.85 and 1.0, the corresponding $Sim_{IC}$ similarities have values between 0.7 and 0.9. Figure 3 C shows an example of different levels of correction by relevance and $Sim_{IC}$ similarities for the MICA with low probability. Yeast proteins YDR130C and YOR058C both annotated with term "microtubule binding", a level-six GO term with 10 annotations in total associated with it and its descendants. Yeast proteins YDR130C and YOR058C will have similarity values of 1.0 based on Lin, Wang, Jiang and GIC measures. The relevance similarity has a value of 0.9984 and the $Sim_{IC}$ has a value of 0.8658.

**Improvement of correlation between GO semantic similarities and sequence similarities of yeast proteins by $Sim_{IC}$**

In order to further demonstrate the effectiveness of $Sim_{IC}$, we first calculated the correlation between GO semantic similarities and sequence similarities of yeast proteins. Ten GO semantic similarities including the Lin, Jiang, GIC, RSS, Resnik and adjusted Resnik as well as two types of relevance and $Sim_{IC}$ similarities were used for analysis. Previous studies [4, 5, 24] already showed that high sequence similarities are correlated with high semantic similarities although high semantic similarities do not always mean high sequence similarities. Therefore, we only employed protein pairs with both significant sequence similarity and semantic similarity in the comparison (see detail in Methods). As a protein can be annotated by multiple GO terms, three approaches have been used to calculate protein semantic similarity from the GO term similarity. Namely, the semantic similarity between proteins is the maximum, average and best match average of GO term similarities. Table 1, 2 and 3 show the correlations between protein semantic similarities based on twelve GO similarity measures of three GO branches and the sequence similarity for each of these three protein similarity approaches. For the average protein similarity approach, the Resnik similarity gave higher average correlation than those original Lin, Jiang gave, which is consistent with Lord's results [5]. For all three approaches of protein semantic similarities, the $Sim_{IC}$ always give higher average correlation values and lower standard deviation than those of Lin, Jiang and GIC similarities. However, the relevance similarities do not always lead to higher average correlations. This indicated that the $Sim_{IC}$ will be more effective than relevance similarity. Furthermore, it is noted that the $Sim_{IC}$ has dramatically improved the correlations between sequence similarities and GO semantic similarities of cellular components (CC), which usually have higher shallow annotation rate. Moreover, the adjusted Resnik similarity gave the highest average correlation for maximum protein similarity approach, and the $Sim_{IC}$ resulted in the highest average correlations for both average and best match average protein similarity approaches. The best match average approach based on $Sim_{IC}$ give overall best average correlation value of 0.893. The scatter plots of best match average based on BP (Figure 5), CC (Figure 6) and MF (Figure 7) $Sim_{IC}$ measured against corresponding sequence similarities showed a strong linear relationship. These results suggest that $Sim_{IC}$ measure can improve the average correlation between semantic similarity and sequence similarity of proteins.

**Evaluation of semantic similarities of protein pairs with various levels of sequences similarities**

To further test the correlation between semantic and sequences similarities, we calculated the distribution of the protein BP and MF semantic similarity of proteins pairs in different levels of evolutionary relationship [24]. The protein pairs between human and yeast were divided into four categories based on the level of sequence similarity: orthology (from Inparanoid database [30]), high similarity (HS), low similarity (LS) and no similarity (NS). The distributions of BP and MF $Sim_{IC}$ similarities in four categories are shown in Figure 8 and 9. The NS dataset has the highest percentage of low semantic similarity and lowest percentage of high semantic similarity. On the other hand, the orthology data has the lowest percentage of low semantic similarity and highest percentage of high semantic similarity. Figure 9 shows a histogram of the relationship between MF and BP semantic similarity for the protein pairs in the orthology dataset. Similar to results of Schlicker et al., the highest peak occurs at M0.9 and B0.9. Next, we compared our distributions with distributions of GIC (Figures 11, 12) and relevance similarity (Figures 13, 14) in four datasets. The distributions of BP GIC similarities show a higher percentage in the low semantic similarity range and a lower percentage in the high semantic similarity range A Chi-square test showed that the distributions of BP $Sim_{IC}$ similarities are significant different from those of GIC similarity with p-value less than $10^{-7}$ in all four datasets. The distribution of MF $Sim_{IC}$ similarities are also significant different from those of GIC similarity with p-value less than 0.025, except for NS dataset. Meanwhile, the distributions of BP and MF relevance similarities are very similar to those of $Sim_{IC}$ similarities. A Chi-square test showed no significant differences with p-values greater than 0.75.

**Improvement of correlation between GO semantic similarities and expression similarities of yeast proteins by $Sim_{IC}$**

Another widely used method to evaluate GO semantic similarities is to calculate the correlation between semantic similarities and gene expression similarities [7]. Here, we calculated the correlations between each of these three approaches of protein semantic similarities based on the Lin, Jiang, GIC, RSS, Resnik and adjusted Resnik as well as two types of relevance and $Sim_{IC}$ similarities of three GO branches and each of the expression similarities from four yeast microarray studies [31-34]. The results are shown in Tables 4, 5 and 6. For the maximum protein similarity approach, the Resnik similarity yielded higher average correlations than Lin and Jiang similarities gave, which is consistent with Sevilla's results [7]. For all these three approaches of protein semantic similarities, the $Sim_{IC}$ always gave higher average correlation values than those Lin, Jiang and GIC similarities give, although the standard deviation of correlations are not always lower. Meanwhile, the relevance similarities only gave lower average correlations than those Lin, Jiang and GIC similarities gave in some of cases. These results further confirm that the $Sim_{IC}$ will be more effective than relevance similarity in fine tuning up the semantic similarities. The Resnik similarity resulted in the highest average correlation for maximum protein similarity approach and the $Sim_{IC}$ yielded the highest average correlations for both average and best match average protein similarity approaches. Overall, the best match average approach based on $Sim_{IC}$ gave the highest average correlation value of 0.785. These results suggest that $Sim_{IC}$ will improve the average correlations between semantics similarities and gene expression similarities of proteins.

**Improvement of protein interaction prediction using $Sim_{IC}$**

Recently, Wu et al. [28] predicted protein interactions based on RSS between protein pairs. They predicted "gold standard" protein interactions using 0.8 as threshold for BP and CC RSS similarity values. Their predictions are proven to be consistent with known experimental interactions and complexes. RSS is a similarity measure based on the structural information only. As discussed in the introduction above, it may be insufficient to measure the similarity between protein pairs. So we hypothesized that predicting protein interactions based on RSS may not be effective. In an example provided in Wu's paper, the authors proposed that LRP1 is a new member of complex DNA ligase IV (MIPS complex 510.180.30.20) because LRP1 has high RSS similarities with other two member of this complex, DNL4 and LIF1. For the maximum protein similarity approach, the RSS between DNL4 and LRP1and between LRP1 and LIF1 are 0.818 for CC and 1.000 for BP. Meanwhile, for the maximum protein similarity approach, the $Sim_{IC}$ similarities between DNL4 and LRP1and between LRP1 and LIF1 are both 0.194 for CC and 0.864 for BP. By looking into cellular component GO terms annotated LRP1, DNL4 and LIF1, we find that the MICA of CC GO terms annotated LRP1 and LIF1 is "nuclear part" with id GO:0044428 (see Figure 15), also inferred by the MICA of CC GO terms annotated LRP1 and DNL4 (see Figure 16). Although the "nuclear part" is at the fifth level of the GO cellular component tree, there are 1662 annotations in total associated with this GO term and its descendants, which means that this GO term is a general GO term and is weakly informative. Our $Sim_{IC}$ considers this fact, hence providing lower similarity values.

We also observed that some protein pairs with low RSS actually score quite high with $Sim_{IC}$ similarities. An example is yeast protein pair HSP12 and MUC1. The BP RSS of this pair is 0.270 while the $Sim_{IC}$ is 0.847. After further investigating the BP GO terms annotated HSP12 and MUC1, we found that the MICA (also the most recent common ancestor (MRCA)) of these BP GO terms is "cell adhesion" with id GO:0007155 (see Figure 17). Although the "cell adhesion" is at the third level of GO biological process tree, there are only 9 annotations associated with this GO term and its descendants, which means this term is very informative. Our $Sim_{IC}$ similarities take into account both structure and information content, and then give out high similarity value between HSP12 and MUC1. On the other hand, RSS gives low similarity value purely based on the structure information of the GO terms. By integrating the information content and structural information of the GO, our $Sim_{IC}$ similarities should be more effective in predicting protein interactions.

To further investigate the effect of information content on RSS measure, we compared the RSS values of all possible yeast protein pairs with those of $Sim_{IC}$. As Wu et al. used the RSS values 0.8 as threshold to determine their gold standard positive, we wanted to examine how many protein pairs with high RSS values (>0.8) will have a low $Sim_{IC}$ (<=0.2). A statistical study (see Methods) showed that protein pairs were unlikely to interact if their $Sim_{IC}$ is less than or equal to 0.2. As shown in Table 7, we obtained 234976 protein pairs with high BP RSS (>0.8) (see Supplementary Table 1). Among these 234976 protein pairs, 681 pairs have low $Sim_{IC}$ similarities (<=0.2) with a percentage of 0.29%. We also obtained 172862 pairs of proteins with high CC RSS (>0.8) (see Supplementary Table 2). Among these, 81515 pairs had low $Sim_{IC}$ values (<=0.2) with a percentage of 5.96%. Furthermore, for 12956 pairs of proteins with both BP and CC RSS values greater than 0.8 (see supplementary table 3), 1460 pairs with either BP or CC $Sim_{IC}$ values are less than or equal 0.2 with a percentage of 11.27%. We believe the reason that we detected fewer pairs of interacting proteins than those obtained by Wu et al. may be because of the different version of GO annotation we used or because of only "is_a" and "part_of" relationship in DAG tested in this study. Our results show that "shallow annotation" affects the calculation of semantics similarity. RSS, which utilize only the structural information of GO DAG, may lead to a ~10% increase in false positive protein interaction predictions.

Then, we applied our $Sim_{IC}$ to predict yeast protein interactions. In order to compare with Wu's study, we used the maximum protein similarity approach. First, the statistical significances of $Sim_{IC}$ were obtained by comparing the $Sim_{IC}$ similarities of interactions between proteins from MIPS complex [35, 36] with those of randomly sampled protein interactions (see Methods section for details). As shown in Figure 18, in two of 100 bins: BP (0.7, 0.8], CC (0.8, 0.9] and BP (0.8, 0.9], CC (0.8, 0.9], the MIPS complex protein interactions were 1580 and 774 standard deviation greater than the average number of random protein pair in those bins, respectively. Following Wu et al, we use 0.7 as threshold for the BP $Sim_{IC}$ and 0.8 as threshold for CC $Sim_{IC}$ to predict yeast protein interactions. If a protein pair's BP $Sim_{IC}$ was greater than 0.7 and its CC $Sim_{IC}$ was greater than 0.8, they were predicted to interact. In total, we predicted 20484 protein-protein interactions (PPIs) between 2462 proteins (Table 8 and Supplemental Table 4). Among 20484 PPIs, there were 6076 PPIs are in the MIPS data, which includes 8250 protein pairs with both BP and CC annotations. Thus, our predictions revealed 73.65% of MIPS PPIs

Figure 19 shows the predicted protein interaction network. The network consists of 128 connected components. The largest component has 1871 proteins and 18269 interactions. As an example, we examined the predicted interactions of two proteins mentioned above in complex DNA ligase IV: DNL4 and LIF1. Figure 20 shows that, in predicted PPIs, the DNL4 and LIF1 interact and both are connected to two proteins in the Ku complex: YKU80 and YKU70. The link between DNL4 and LIF1 and the link between YKU70 and YKU80 are verified in both MIPS database [35] and DIP database [37]. The links between DNL4, LIF1 and YKU70, YKU80 are not presented in both databases. However, there is experimental evidence showed that Ku complex can form a temporary complex with DNA ligase IV complex [38]. Then, we examined how MIPS complexes [35] exist in our predicted PPIs. Only 214 MIPS complexes with at least two distinct members were included in our study. Members of 202 complexes were found in our predicted PPIs (See Supplemental Table 5). There were 159 complexes of which all members were detected. Meanwhile, only 120 of 214 complexes of which all members were found in Wu's predicted protein network [28]. This result implies our predicted networks are more consistent in detecting experimentally verified protein complexes.

**Assessment of prediction of protein-protein interactions using different similarity measures**

To further examine the effectiveness of $Sim_{IC}$ in PPI prediction, we used the area under receiver operating characteristics (ROC) curve (AUC) [39] method to evaluate the performance of classifiers based on different semantic similarity measures. The 8250 protein pairs from MIPS [35, 36] complex data with both BP and CC annotation were used as positive controls and an identical number of protein pairs were selected randomly as a negative control dataset. Three types of protein similarity approach based on Lin, Jiang, Resnik, GIC, relevance, $Sim_{IC}$, and RSS were tested and the combination of BP and CC similarities were used to predict PPIs. Tests were repeated ten times with different sampling negative control dataset have been performed. As shown in Table 9, although the performances of relevance, $Sim_{IC}$ and GIC semantic measures were close, the best match average approach based on $Sim_{IC}$ give overall best performance as indicated by the AUC value. This comparison demonstrates that the proposed $Sim_{IC}$ is superior in predicting yeast protein interactions.

## Conclusions

The Gene Ontology has been increasingly used to annotate proteins, and in some cases is the only clue to biological function in a newly sequenced gene. The effectiveness of semantic similarity measures is an important consideration for the comparison of proteins based solely on their GO annotations. Most current existing semantic similarity measures do not effectively combine the structural information of GO DAG and information content of GO terms. In this paper, we proposed a new measure, $Sim_{IC}$, which combines both information content of GO terms and structural information of Gene Ontology hierarchy. The $Sim_{IC}$ can efficiently address the shallow annotation problem. We compared our proposed $Sim_{IC}$ measure with Lin, Jiang, GIC, relevance, RSS, Resnik and adjusted Resnik similarities through the study of the correlation between semantic similarities of yeast protein pairs and their gene expression similarities and the correlation between semantic similarities of yeast protein pairs and their sequence similarities. The results show that the similarities obtained by our $Sim_{IC}$ measure have the highest average correlation with gene expression similarities and sequence similarities. Furthermore, we demonstrate that our $Sim_{IC}$ measure is superior to RSS in predicting protein interactions. We have successfully predicted 20484 protein-protein interactions (PPIs) between 2462 proteins, and the predicted protein interaction network covers more MIPS protein complex that the protein interaction network of Wu et al. does. Moreover, a quick examination of our PPI network detected the predicted protein interactions between Ku complex and DNA ligase IV complex, which are supported by biological evidence.

Although the $Sim_{IC}$ measure is promising, there are reasons to further improve the semantic similarity measure. In our current approach, we have not distinguished the "is_a" and the "part_of" relationships among GO terms. In the future, we will investigate the difference of the information content between the "is_a" and the "part_of" relationships. In addition, two terms are semantically close if they have "regulates", "positively_regulates" and "negatively_regulates" relation with the same term, even though they are not directly related. Utilization of the "regulates", "positively_regulates" and "negatively_regulates" relationships in the semantic similarity measure will require further investigation. Furthermore, it is worthwhile to explore the role of evidence codes of GO annotation in the semantic similarity measure.

## Methods
### GO database and Yeast GO annotation

The Gene Ontology dated April 16th 2008 was downloaded from gene ontology consortium website (www.geneontology.org). The yeast gene annotation database was from Saccharomyces Genome Database ([www.yeastgenome.org](www.yeastgenome.org)) and the April 12th 2008 release was used. Proteins only annotated with "biological process", "molecular function" and "cellular component" were filtered out from the analyses. And the annotations with IEA evidence code are also ignored in this study. After this filtering, there were 4620 proteins with biological process annotations, 3876 proteins with molecular function annotation and 5077 proteins with cellular component annotation. Furthermore, only the "is_a" and "part_of" relationship are used to build the GO DAG. The "regulates", "positively_regulates" and "negatively_regulates" relationships are not included in our present study.

**New semantic similarity measures by integrating IC and structure of GO DAG**

**Information coefficient similarity (Sim$_{IC}$)**. The Sim$_{IC}$ between two GO term $t_1$ and $t_2$ is defined as:

$$Sim_{IC}(t_1,t_2) = \frac{2 \times \log(prob(MICA))}{\log(prob(t_1)) + \log(prob(t_2))} \times (1 - \frac{1}{1 + IC(MICA)}) \tag{3}$$

Theoretically, the values of Sim$_{IC}$ are also between zero and one. Practically, the yeast has only about 13683 BP annotations, and then the highest IC of BP GO term will be about 9.52 on natural log base. Therefore, the BP Sim$_{IC}$ will have maximum values around 0.9.

**Adjusted Resnik similarity**. In order to add structural information into Resnik similarity, we simply adjust it with Lin similarity:

$$Sim_{Adjusted-Resnik} = Sim_{Resnik} \times Sim_{Lin} \tag{4}$$

Similar to original Resnik similarity, this adjusted Resnik similarity has no maximum value and has a minimum zero.

**Existing semantic similarity measures**

**Resnik similarity**. Resnik introduced a semantic similarity for "is_a" ontologies based the highest IC values among IC values of all common ancestors of two terms [21]:

$$Sim_{Resnik}(t_1,t_2) = \max_{t \in S(t_1,t_2)}(IC(t)) \tag{5}$$

where $S(t_1, t_2)$ is the set of common ancestors of two term $t_1$ and $t_2$. The Resnik similarity has no maximum values and has a minimum zero.

**Lin similarity**. Lin [23] developed a information-theoretic similarity applicable to any domain that can be described by a probabilistic model. Lin measure is based on the relative probability between two terms and their MICA. The Lin measure between two GO term $t_1$ and $t_2$ is defined as:

$$Sim_{Lin}(t_1,t_2) = \frac{2 \times \log(prob(MICA))}{\log(prob(t_1)) + \log(prob(t_2))} \tag{6}$$

The value of Lin similarity ranges from zero to one.

**Jiang similarity.** The Jiang and Conrath [22] integrated the edge-based method with the node-based approach of the information content calculation to develop a new distance measure. For its simple case in which factors related to local density, node depth and link type are ignored, the Jiang measure between two GO term $t_1$ and $t_2$ is defined as:

$$Dis_{Jiang}(t_1, t_2) = IC(t_1) + IC(t_2) - 2 \times IC(MICA(t_1, t_2)) \quad (7)$$

Jiang distance measure can easily be transformed into a similarity measure by adding one and inverting it [10].

$$Sim_{Jiang}(t_1, t_2) = \frac{1}{1 + Dis_{J-C}(t_1, t_2)} \quad (8)$$

If terms t1 and t2 are the same, $Dist_{Jiang}(t1, t2)$ should be 0. Adding one is to avoid the division of 0. The value of Jiang similarity ranges from zero to one

**Graph information content (GIC) similarity**. Let DAGT1 and DAGT2 be two ancestor DAGs induced by two GO terms $t_1$ and $t_2$ [21]. Then, the graph information content (GIC) [26] measure is defined as the ratio between sum of information content of GO terms in the intersection of DAGT1 and DAGT2 and sum of information content of GO terms in the union of DAGT1 and DAGT2:

$$Sim_{GIC}(t_1, t_2) = \frac{\sum_{t \in DAG^{T1} \cap DAG^{T2}} IC(t)}{\sum_{t \in DAG^{T1} \cup DAG^{T2}} IC(t)} \quad (9)$$

The values of GIC similarity ranges from zero to one.

**Relative Specificity Similarity (RSS)**. The RSS of two GO terms is obtained from three components α, β and γ. Component α is length of path from the root to most recent common ancestor (MRCA). Component β is the maximum length of paths from each of GO terms to its leaf descendants. Component γ is the sum of distance between MRCA and two GO terms. Thus, the RSS between two GO terms can be calculated by[28]:

$$RSS(t_1, t_2) = \frac{\max Depth^{GO}}{\max Depth^{GO} + \gamma} + \frac{\alpha}{\alpha + \beta} \quad (10)$$

where the maxDepthGO is the maximum distance from the GO root to the leaf term. The values of RSS are between 0 and 1.

**Schlicker's relevance similarity**. Schlicker et al. pointed out that both the specific of MICA and the relation between two GO terms with MICA are needed to be considered in a similarity measure. Then, they developed a new relevance similarity measure by combining the Lin similarity with the probability of MICA [24].

$$Sim_{Rel}(t_1,t_2) = \max_{t \in S(t_1,t_2)} (Sim_{Lin} \cdot (1 - prob(MICA))) \tag{11}$$

The values of relevance similarity are also between zero and one. We also evaluate the varieties of relevance similarity by using probability of MICA to adjust Jiang similarity.

**Calculation of similarities between proteins**

A protein can be annotated with a set of GO terms. Therefore, similarity between two proteins can be calculated based on the similarities between two set of GO terms. Here, we investigated three approaches for computing protein similarity based on GO term similarities [4, 13, 24, 26].

**Best match average similarity**. The similarity between two proteins is based on average of best match GO term pairs [13]. Given two proteins $p_1$ and $p_2$, let go1 and go2 represent their corresponding sets of GO terms. First, the similarity between one GO term t and a set of GO terms, go = {$t_1$, $t_2$, … $t_k$) is defined as the maximum similarity between the t and any member in set go.

$$Sim(t, go) = \max_{1 \leq i \leq k}(Sim(t, t_i)) \tag{12}$$

Therefore, the similarity between two proteins can be defined as the weighted average of the term similarity scores:

$$Sim(p_1, p_2) = \frac{\sum_{1 \leq i \leq m} Sim(t_{1i}, go_2) + \sum_{1 \leq j \leq n} Sim(t_{2j}, go_1)}{m + n} \tag{13}$$

where m is the number of terms in go1 and n is the number of terms in go2. In the best match average protein similarity measures, proteins with more GO terms annotated will have more influence on the overall similarity score.

**Average similarity.** The similarity between two proteins, $p_1$ and $p_2$, is the average of similarities among all pairs of two GO term sets, go1 and go2:

$$Sim(p_1, p_2) = \frac{\sum_{1 \leq i \leq m, 1 \leq j \leq n} Sim(t_i, t_j)}{m \times n} \qquad (14)$$

where m is the number of terms in go1 and n is the number of terms in go2.

**Maximum similarity.** The similarity between two proteins, $p_1$ and $p_2$, is the maximum similarity among all pairs of two GO term sets, go1 and go2:

$$Sim(p_1, p_2) = \max_{1 \leq i \leq m, 1 \leq j \leq n} (Sim(t_i, t_j)) \qquad (15)$$

where m is the number of terms in go1 and n is the number of terms in go2.

**Calculation of correlation between semantic similarity and sequence similarity**

The correlation between semantic similarity and sequence similarity is measured by the correlation between semantic similarity values and the BLAST log bit scores [4, 5] between sequences. An all-against-all BLAST search for yeast proteins are performed on a local copy of NCBI BLAST [40]. The expect value threshold 100 was used. As the BLAST result is not symmetric, for each protein pair, the final sequence similarity is the average of two BLAST results between them [26].

$$Sim_{seq}(P_1, P_2) = \log_{10}(\frac{BScore(P_1, P_2) + BScore(P_2, P_1)}{2}) \qquad (16)$$

where BScore is BLAST bit score. Then, the correlation is calculated based on protein pairs with both BLAST bit scores and semantic similarity values. First, the semantic similarities were split into 50 intervals. For each interval, the average of semantic similarities that fall in this interval and the average of corresponding BLAST log bit scores are calculated. Therefore, the Pearson correlation is calculated based on these 50 pairs of average values.

**Calculation of correlation between semantic similarity and gene expression similarity**

Four microarray profiles have been used for the calculation [31-34]. All four data sets have at least 50 experiment data points, which should ensure the robust of the calculation of correlation between gene expressions. First, the Pearson correlations between expressions of genes were calculated. Only genes with the number of missing values less than ten percentage of total experiments were included. Missing values were estimated using nearest neighbour based method [41]. Only the protein pairs with both expression correlations value and semantic similarity values were included for analysis. The semantic similarities were split into 50 intervals. Then, the correlation between semantic similarity and expression similarity was calculated based on the average values of intervals [7]. For each interval, the average of semantic similarities that fall in this interval and the average of expression correlations between corresponding protein pairs were calculated. Then, the Pearson correlation is calculated based on the average values of semantic similarities and expression correlation values of 50 intervals.

**Determination of statistical significant $Sim_{IC}$ of yeast protein interactions**

First, the BP and CC SimIC similarities of 8250 interactions between proteins from MIPS complex [35, 36] were calculated. Then, the BP or CC SimIC similarities values were partitioned into 10 bins, respectively. The *i*th bin includes values with i*bin_size <values≤ (i+1)*bin_size, where bin_size=0.1. Combining the BP and CC similarity bins, there were 100 bins in total. The protein pairs were distributed into 100 bins based on the combination of BP and CC similarity values. Thereafter, we randomly sample 8250 protein interactions and calculate their BP and CC SimIC similarities. The number of protein pairs in each of 100 bins was counted based on the BP and CC similarity values. We repeated this process one thousand times. Finally, the number of MIPS complex protein interactions in each bin was compared to the average and standard deviation of number of random protein interactions in each bin to obtain a Z score:

$$Z\ Score = (\#\ paris_{MIPS} - Average\#\ pairs_{random})/SD_{random} \qquad (17)$$

The Figure 18 shows the Z score in each of 100 bins. The MIPS complex protein interactions have shown extremely high Z scores in two of 100 bins: BP (0.7, 0.8], CC (0.8, 0.9] and BP (0.8, 0.9], CC (0.8, 0.9].

**ROC curve analysis**

The receiver operating characteristics (ROC) [39] evaluates the performance of classifiers based on the tradeoff between specificity and sensitivity. The area under the ROC curve (AUC) can be used to compare the prediction performance. While an area of 1 means perfect prediction, an area of 0.5 indicates random prediction. We employed the ROCR [42] library to draw the ROC curves and calculate the AUC values.

# Authors' contributions
FL developed the methods. BL and FL implemented the methods. BL, JZW, AF JZZ and FL wrote the paper.

# Acknowledgements
FL and BL are supported by NSF EPSCoR grants EPS-0447660. Feng Luo would like to acknowledge financial support from the Institute for Modeling and Simulation Applications at Clemson University. The authors would like to thank Dr. Andread Schlicker for providing the IO, HSS, LSS and NSS datasets. JZW' work is partially supported by the National Institutes of Health contract 1R15CA131808-01.

# Figures

**Figure 1 – Gene ontology DAG illustration**

**Figure 2 – Distribution of Lin, relevance and $Sim_{IC}$ of all pairs of annotated yeast proteins**

The values of all similarities are partitioned into 21 bins. The first bin includes number of protein pairs with zero values. The following $i$th bin includes values number of protein pairs with i*bin_size <values≤ (i+1)*bin_size, where bin_size=(max-min)/20. Inner figure shows the distribution of last four bins.

**Figure 3 – Examples of different levels of information coefficient corrections**

**Figure 4 – Scatter plot of relevance similarities against $Sim_{IC}$ similarites of yeast protein pairs with Lin's similarity greater than 0.9**

**Figure 5 – Scatter plot of yeast protein pair BP semantic similarity against their sequence similarity**

The protein similarities are calculated using best math average method and the $Sim_{IC}$ is used to calculate semantic similarities among GO terms. The semantic similarities are split into 50 intervals. For each interval, the average of BP semantic similarities that fall in this interval and the average of corresponding BLAST log bit scores are calculated. The trend line is a linear regression of the data.

**Figure 6 – Scatter plot of yeast protein pair CC semantic similarity against their sequence similarity**

The protein similarities are calculated using best math average method and the $Sim_{IC}$ is used to calculate semantic similarities among GO terms. The semantic similarities are split into 50 intervals. For each interval, the average of CC semantic similarities that fall in this interval and the average of corresponding BLAST log bit scores are calculated. The trend line is a linear regression of the data.

**Figure 7 – Scatter plot of yeast protein pair MF semantic similarity against their sequence similarity**

The protein similarities are calculated using best math average method and the $Sim_{IC}$ is used to calculate semantic similarities among GO terms. The semantic similarities are split into 50 intervals. For each interval, the average of MF semantic similarities that fall in this interval and the average of corresponding BLAST log bit scores are calculated. The trend line is a linear regression of the data.

**Figure 8 – Distribution of the $Sim_{IC}$ BP similarities of human and yeast protein pairs in four different sets**

The bins correspond to the following intervals of protein similarity values: S0.0: [min, min+0.2*range]; S0.2: [min+0.2*range, min+ 0.4*range]; S0.4: [min+0.4*range, min+0.6*range]; S0.6: [min+0.6*range, min+0.8*range]; S0.8: [min+0.8*range, max]. Here, min and max is the minimum and maximum of similarities and range = max-min.

**Figure 9 – Distribution of the $Sim_{IC}$ MF similarities of human and yeast protein pairs in four different sets**

The bins correspond to the same intervals as in legend of Figure 7.

**Figure 10 – Distribution of the $Sim_{IC}$ BP and MF human and yeast protein similarities for orthology dataset**

The bins correspond to the same intervals as in legend of Figure 7.

**Figure 11 – Distribution of the GIC BP similarities of human and yeast protein pairs in four different sets**

The bins correspond to the same intervals as in legend of Figure 7.

**Figure 12 – Distribution of the GIC MF similarities of human and yeast protein pairs in four different sets**

The bins correspond to the same intervals as in legend of Figure 7.

**Figure 13 – Distribution of the relevance BP similarities of human and yeast protein pairs in four different sets**

The bins correspond to the same intervals as in legend of Figure 7.

**Figure 14 – Distribution of the relevance MF similarities of human and yeast protein pairs in four different sets**

The bins correspond to the same intervals as in legend of Figure 7.

**Figure 15 – Visualization of GO annotations for yeast genes LRP1 and LIF1**

The orange nodes are GO terms annotating LRP1 and cyan nodes are GO terms annotating gene LIF1.

**Figure 16 – Visualization of GO annotations for yeast genes LRP1 and DNL4**

The orange nodes are GO terms annotating gene LRP1 and the cyan nodes are GO terms annotating gene DNL4.

**Figure 17 – Visualization of GO annotations for yeast genes HSP12 and MUC1**

The orange nodes are GO terms annotating gene HSP12 and the cyan nodes are GO terms annotating gene MUC1.

**Figure 18 – Statistical significance of SimICsimilarity ($Sim_{IC}$)**

The values of BP $Sim_{IC}$ and CC $Sim_{IC}$ are partitioned into 10 bins, respectively. The *i*th bin includes values number of protein pairs with i*bin_size <values≤ (i+1)*bin_size, where bin_size=0.1. The Z-score for each combination bin of BP and CC are calculated by (#pairs$_{MIPS}$ –Average #pairs$_{random}$)/SD$_{random}$.

**Figure 19 – The predicted yeast protein interaction networks using $Sim_{IC}$**

The graph is draw using cytoscape [43].

**Figure 20 – Predicted protein interactions between Ku complex (YKU70, YKU80) and DNA ligase IV complex (DNL4, LIF1)**

The existing protein interactions are represented by black links. The predicted protein interactions are represented by red links.

# Tables

### Table 1 - Correlations between semantic similarities (maximum) and sequence similarities of yeast proteins

The varieties of relevance similarity and $Sim_{IC}$ also are used in comparison. The highest average correlation is indicated by bold.

|  | No adjustment | | | | | | Relevance similarities | | $Sim_{IC}$ | |
|---|---|---|---|---|---|---|---|---|---|---|
|  | RSS | Resnik | Resnik*lin | Lin | jiang | GIC | lin | Jiang | lin | jiang |
| BP | 0.691 | 0.899 | 0.911 | 0.714 | 0.505 | 0.777 | 0.701 | 0.593 | 0.738 | 0.709 |
| MF | 0.694 | 0.905 | 0.894 | 0.839 | 0.311 | 0.705 | 0.879 | 0.544 | 0.855 | 0.765 |
| CC | 0.535 | 0.861 | 0.870 | 0.732 | 0.465 | 0.702 | 0.657 | 0.558 | 0.767 | 0.738 |
| Average | 0.640 | 0.888 | **0.892** | 0.762 | 0.427 | 0.728 | 0.746 | 0.565 | 0.787 | 0.737 |
| Std | 0.091 | 0.024 | 0.021 | 0.068 | 0.102 | 0.042 | 0.118 | 0.025 | 0.061 | 0.028 |

### Table 2 - Correlations between semantic similarities (average) and sequence similarities of yeast proteins

The varieties of relevance similarity and $Sim_{IC}$ also are used in comparison. The highest average correlation is indicated by bold.

|  | No adjustment | | | | | | Relevance similarities | | $Sim_{IC}$ | |
|---|---|---|---|---|---|---|---|---|---|---|
|  | RSS | Resnik | Resnik*lin | lin | jiang | GIC | lin | Jiang | lin | jiang |
| BP | 0.788 | 0.895 | 0.894 | 0.833 | 0.649 | 0.777 | 0.829 | 0.841 | 0.863 | 0.798 |
| MF | 0.803 | 0.898 | 0.856 | 0.906 | 0.687 | 0.705 | 0.868 | 0.623 | 0.872 | 0.656 |
| CC | 0.631 | 0.616 | 0.460 | 0.570 | 0.361 | 0.702 | 0.664 | 0.681 | 0.735 | 0.774 |
| Average | 0.741 | 0.803 | 0.737 | 0.770 | 0.566 | 0.728 | 0.787 | 0.715 | **0.823** | 0.743 |
| Std | 0.095 | 0.162 | 0.240 | 0.177 | 0.178 | 0.042 | 0.108 | 0.113 | 0.077 | 0.076 |

### Table 3 - Correlations between semantic similarities (best match average) and sequence similarities of yeast proteins

The varieties of relevance similarity and $Sim_{IC}$ also are used in comparison. The highest average correlation is indicated by bold.

|  | No adjustment | | | | | | Relevance similarities | | $Sim_{IC}$ | |
|---|---|---|---|---|---|---|---|---|---|---|
|  | RSS | Resnik | Resnik*lin | lin | jiang | GIC | lin | Jiang | lin | jiang |
| BP | 0.825 | 0.931 | 0.922 | 0.910 | 0.889 | 0.777 | 0.895 | 0.908 | 0.915 | 0.935 |
| MF | 0.832 | 0.942 | 0.910 | 0.848 | 0.754 | 0.705 | 0.938 | 0.710 | 0.892 | 0.838 |
| CC | 0.714 | 0.647 | 0.545 | 0.770 | 0.759 | 0.702 | 0.840 | 0.789 | 0.871 | 0.826 |
| Average | 0.790 | 0.840 | 0.792 | 0.843 | 0.801 | 0.728 | 0.891 | 0.802 | **0.893** | 0.866 |
| Std | 0.066 | 0.167 | 0.214 | 0.070 | 0.077 | 0.042 | 0.049 | 0.100 | 0.022 | 0.060 |

**Table 4 - Correlations between semantic similarities (maximum) and expression similarities of yeast proteins**

The varieties of relevance similarity and $Sim_{IC}$ also are used in comparison. The highest average correlation is indicated by bold. Four microarray profiles [31-34] are denoted as "MAPK" [33], "Damage"[31], "cell cycle"[34] and "ENV" [32], respectively. Noted that the GIC semantic similarity measure itself is a protein similarity measure and has the same values in Table 4, 5, 6.

|  | No adjustment | | | | | | Relevance similarities | | $Sim_{IC}$ | |
|---|---|---|---|---|---|---|---|---|---|---|
|  | RSS | Resnik | Resnik*lin | lin | jiang | GIC | lin | Jiang | lin | jiang |
| BP-MAPK | 0.511 | 0.631 | 0.575 | 0.575 | 0.200 | 0.774 | 0.499 | 0.354 | 0.568 | 0.378 |
| MF-MAPK | 0.419 | 0.564 | 0.638 | 0.505 | 0.007 | 0.539 | 0.556 | 0.147 | 0.640 | 0.532 |
| CC-MAPK | 0.583 | 0.550 | 0.62 | 0.323 | 0.209 | 0.530 | 0.299 | 0.392 | 0.519 | 0.372 |
| BP-Damage | 0.497 | 0.849 | 0.715 | 0.727 | 0.398 | 0.716 | 0.568 | 0.505 | 0.716 | 0.531 |
| MF-Damage | 0.464 | 0.752 | 0.800 | 0.682 | 0.275 | 0.611 | 0.533 | 0.479 | 0.707 | 0.669 |
| CC-Damage | 0.672 | 0.640 | 0.671 | 0.472 | 0.043 | 0.666 | 0.477 | 0.205 | 0.630 | 0.469 |
| BP-cell cycle | 0.422 | 0.894 | 0.733 | 0.760 | 0.501 | 0.793 | 0.583 | 0.528 | 0.756 | 0.580 |
| MF-cell cycle | 0.537 | 0.844 | 0.838 | 0.615 | 0.022 | 0.636 | 0.581 | 0.455 | 0.774 | 0.746 |
| CC-cell cycle | 0.668 | 0.806 | 0.824 | 0.699 | 0.616 | 0.721 | 0.687 | 0.393 | 0.798 | 0.697 |
| BP-ENV | 0.427 | 0.884 | 0.797 | 0.731 | 0.466 | 0.757 | 0.612 | 0.527 | 0.731 | 0.547 |
| MF-ENV | 0.36 | 0.786 | 0.791 | 0.541 | 0.290 | 0.604 | 0.543 | 0.348 | 0.737 | 0.681 |
| CC-ENV | 0.707 | 0.718 | 0.733 | 0.398 | 0.040 | 0.658 | 0.484 | 0.295 | 0.686 | 0.520 |
| Average | 0.522 | **0.743** | 0.728 | 0.586 | 0.252 | 0.667 | 0.535 | 0.386 | 0.689 | 0.560 |
| Std | 0.114 | 0.122 | 0.086 | 0.141 | 0.211 | 0.09 | 0.094 | 0.124 | 0.084 | 0.121 |

**Table 5 - Correlations between semantic similarities (average) and expression similarities of yeast proteins**

The varieties of relevance similarity and $Sim_{IC}$ also are used in comparison. The highest average correlation is indicated by bold. Four microarray profiles [31-34] are denoted as "MAPK" [33], "Damage"[31], "cell cycle"[34] and "ENV" [32], respectively. Noted that the GIC semantic similarity measure itself is a protein similarity measure and has the same values in Table 4, 5, 6.

|  | No adjustment | | | | | | Relevance similarities | | $Sim_{IC}$ | |
|---|---|---|---|---|---|---|---|---|---|---|
|  | RSS | Resnik | Resnik*lin | lin | jiang | GIC | lin | Jiang | lin | jiang |
| BP-MAPK | 0.658 | 0.580 | 0.435 | 0.762 | 0.497 | 0.774 | 0.788 | 0.450 | 0.734 | 0.444 |
| MF-MAPK | 0.392 | 0.614 | 0.633 | 0.531 | 0.372 | 0.539 | 0.507 | 0.300 | 0.631 | 0.552 |
| CC-MAPK | 0.570 | 0.321 | 0.254 | 0.426 | 0.185 | 0.530 | 0.384 | 0.423 | 0.462 | 0.355 |
| BP-Damage | 0.641 | 0.803 | 0.640 | 0.794 | 0.657 | 0.716 | 0.832 | 0.593 | 0.832 | 0.558 |
| MF-Damage | 0.577 | 0.782 | 0.732 | 0.704 | 0.358 | 0.611 | 0.578 | 0.581 | 0.746 | 0.705 |
| CC-Damage | 0.534 | 0.516 | 0.420 | 0.610 | 0.517 | 0.666 | 0.530 | 0.565 | 0.602 | 0.529 |
| BP-cell cycle | 0.682 | 0.822 | 0.707 | 0.820 | 0.578 | 0.793 | 0.881 | 0.661 | 0.889 | 0.647 |
| MF-cell cycle | 0.604 | 0.882 | 0.856 | 0.768 | 0.433 | 0.636 | 0.696 | 0.553 | 0.812 | 0.770 |
| CC-cell cycle | 0.728 | 0.529 | 0.387 | 0.735 | 0.629 | 0.721 | 0.708 | 0.606 | 0.661 | 0.564 |
| BP-ENV | 0.695 | 0.794 | 0.670 | 0.789 | 0.597 | 0.757 | 0.836 | 0.631 | 0.870 | 0.562 |
| MF-ENV | 0.649 | 0.822 | 0.768 | 0.660 | 0.122 | 0.604 | 0.595 | 0.561 | 0.751 | 0.715 |
| CC-ENV | 0.632 | 0.544 | 0.456 | 0.608 | 0.526 | 0.658 | 0.568 | 0.569 | 0.650 | 0.553 |
| Average | 0.613 | 0.667 | 0.580 | 0.684 | 0.456 | 0.667 | 0.659 | 0.541 | **0.720** | 0.580 |
| Std | 0.089 | 0.173 | 0.184 | 0.120 | 0.170 | 0.09 | 0.155 | 0.101 | 0.125 | 0.116 |

**Table 6 - Correlations between semantic similarities (best match average) and expression similarities of yeast proteins**

The varieties of relevance similarity and $Sim_{IC}$ also are used in comparison. The highest average correlation is indicated by bold. Four microarray profiles [31-34] are denoted as "MAPK" [33], "Damage"[31], "cell cycle"[34] and "ENV" [32], respectively. Noted that the GIC semantic similarity measure itself is a protein similarity measure and has the same values in Table 4, 5, 6.

|  | No adjustment | | | | | | Relevance similarities | | $Sim_{IC}$ | |
| --- | --- | --- | --- | --- | --- | --- | --- | --- | --- | --- |
|  | RSS | Resnik | Resnik*lin | lin | jiang | GIC | lin | Jiang | lin | jiang |
| BP-MAPK | 0.835 | 0.686 | 0.582 | 0.930 | 0.883 | 0.774 | 0.938 | 0.888 | 0.847 | 0.688 |
| MF-MAPK | 0.46 | 0.690 | 0.656 | 0.622 | 0.661 | 0.539 | 0.551 | 0.534 | 0.665 | 0.632 |
| CC-MAPK | 0.688 | 0.389 | 0.351 | 0.540 | 0.426 | 0.530 | 0.599 | 0.588 | 0.562 | 0.590 |
| BP-Damage | 0.82 | 0.870 | 0.808 | 0.904 | 0.878 | 0.716 | 0.926 | 0.893 | 0.897 | 0.774 |
| MF-Damage | 0.677 | 0.820 | 0.816 | 0.750 | 0.649 | 0.611 | 0.613 | 0.640 | 0.767 | 0.754 |
| CC-Damage | 0.703 | 0.587 | 0.537 | 0.711 | 0.576 | 0.666 | 0.685 | 0.706 | 0.696 | 0.727 |
| BP-cell cycle | 0.855 | 0.792 | 0.721 | 0.900 | 0.884 | 0.793 | 0.922 | 0.913 | 0.942 | 0.836 |
| MF-cell cycle | 0.683 | 0.666 | 0.618 | 0.752 | 0.651 | 0.636 | 0.743 | 0.666 | 0.831 | 0.840 |
| CC-cell cycle | 0.708 | 0.637 | 0.554 | 0.732 | 0.600 | 0.721 | 0.819 | 0.806 | 0.771 | 0.765 |
| BP-ENV | 0.847 | 0.863 | 0.798 | 0.908 | 0.887 | 0.757 | 0.921 | 0.896 | 0.919 | 0.789 |
| MF-ENV | 0.638 | 0.841 | 0.813 | 0.686 | 0.693 | 0.604 | 0.647 | 0.685 | 0.787 | 0.802 |
| CC-ENV | 0.835 | 0.686 | 0.582 | 0.930 | 0.883 | 0.658 | 0.938 | 0.888 | 0.738 | 0.761 |
| Average | 0.46 | 0.690 | 0.656 | 0.622 | 0.661 | 0.667 | 0.551 | 0.534 | **0.785** | 0.747 |
| Std | 0.688 | 0.389 | 0.351 | 0.540 | 0.426 | 0.09 | 0.599 | 0.588 | 0.111 | 0.077 |

**Table 7 - Protein pairs with high RSS values and low $Sim_{IC}$ values**

|  | Number of protein pairs with RSS>0.8 | Number of protein pairs with RSS>0.8 and $Sim_{IC}$<= 0.2 | Percentage |
| --- | --- | --- | --- |
| BP | 234976 | 681 | 0.29% |
| CC | 172862 | 81515 | 5.96% |
| BP-CC | 12956 | 1460 | 11.27% |

### Table 8  Proteins pairs with Sim$_{IC}$ greater than threshold

Number of protein pairs with BP Sim$_{IC}$ >0.7 are listed in first row and number of protein pairs with CC Sim$_{IC}$ >0.8 are listed in second row. Protein pairs with BP Sim$_{IC}$ >0.7 and CC Sim$_{IC}$>0.8 are predicted to be physically interacted (third row). The MIPS dataset include 8250 protein pairs with both BP and CC annotations are used to evaluate recall of prediction.

|        | Number of protein | Number of protein pairs with Sim$_{IC}$ greater than threshold | Number (percentage) of predicted protein pairs in MIPS dataset |
|--------|---|---|---|
| BP     | 4775 | 520506 (4.57%) | 6629 (80.35%) |
| CC     | 2698 | 58478 (0.47%)  | 6502 (78.81%) |
| BP-CC  | 2462 | 20484 (0.08%)  | 6076 (73.65%) |

### Table 9 – Comparison of AUC values using different BP and CC similarities to predict protein- protein interactions

The highest average correlation is indicated by bold. The data showed is the average (Ave) and standard deviation (Std) of ten times of tests with different sampling negative control dataset.

| Protein Similarity measures | Semantic Similarity measures | Resnik | Lin | Jiang | GIC | Relevance | Sim$_{IC}$ | RSS |
|---|---|---|---|---|---|---|---|---|
| BMA | Ave±Std | 0.9586 ±0.007 | 0.9526 ±0.008 | 0.9372 ±0.0011 | 0.9566 ±0.0006 | 0.9655 ±0.0004 | **0.9664 ±0.0004** | 0.9316 ±0.0006 |
| MAX | Ave±Std | 0.9448 ±0.0014 | 0.9252 ±0.0012 | 0.9161 ±0.0011 | 0.9566 ±0.0006 | 0.9637 ±0.0006 | 0.9637 ±0.0006 | 0.9479 ±0.0006 |
| Average | Ave±Std | 0.9480 ±0.0008 | 0.9391 ±0.0011 | 0.9224 ±0.0013 | 0.9566 ±0.0006 | 0.9541 ±0.0008 | 0.9553 ±0.0007 | 0.8726 ±0.0008 |

## Additional files

**Additional file 1 – Supplemental Table S1**

Protein pairs with high BP RSS values (>0.8) and low BP $Sim_{IC}$ values (<=0.2).

**Additional file 2 – Supplemental Table S2**

Protein pairs with high CC RSS values (>0.8) and low CC $Sim_{IC}$ values (<=0.2).

**Additional file 3 – Supplemental Table S3**

Protein pairs with both high BP RSS values (>0.8) and high CC RSS values (>0.8) and either low BP or low CC $Sim_{IC}$ values (<=0.2).

**Additional file 4 – Supplemental Table S4**

Protein pairs with both high BP $Sim_{IC}$ values (>0.7) and CC $Sim_{IC}$ values (>0.8), which is predicted as true protein interactions.

**Additional file 5 – Supplemental Table S5**

Members of MIPS protein complexes exist in predicted protein networks.

Figure 1

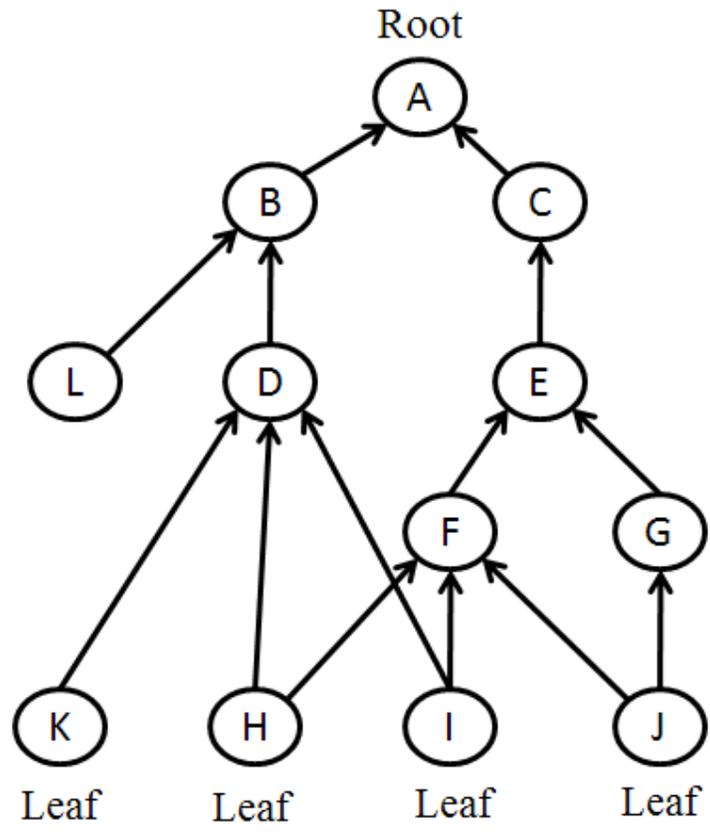

Figure 2

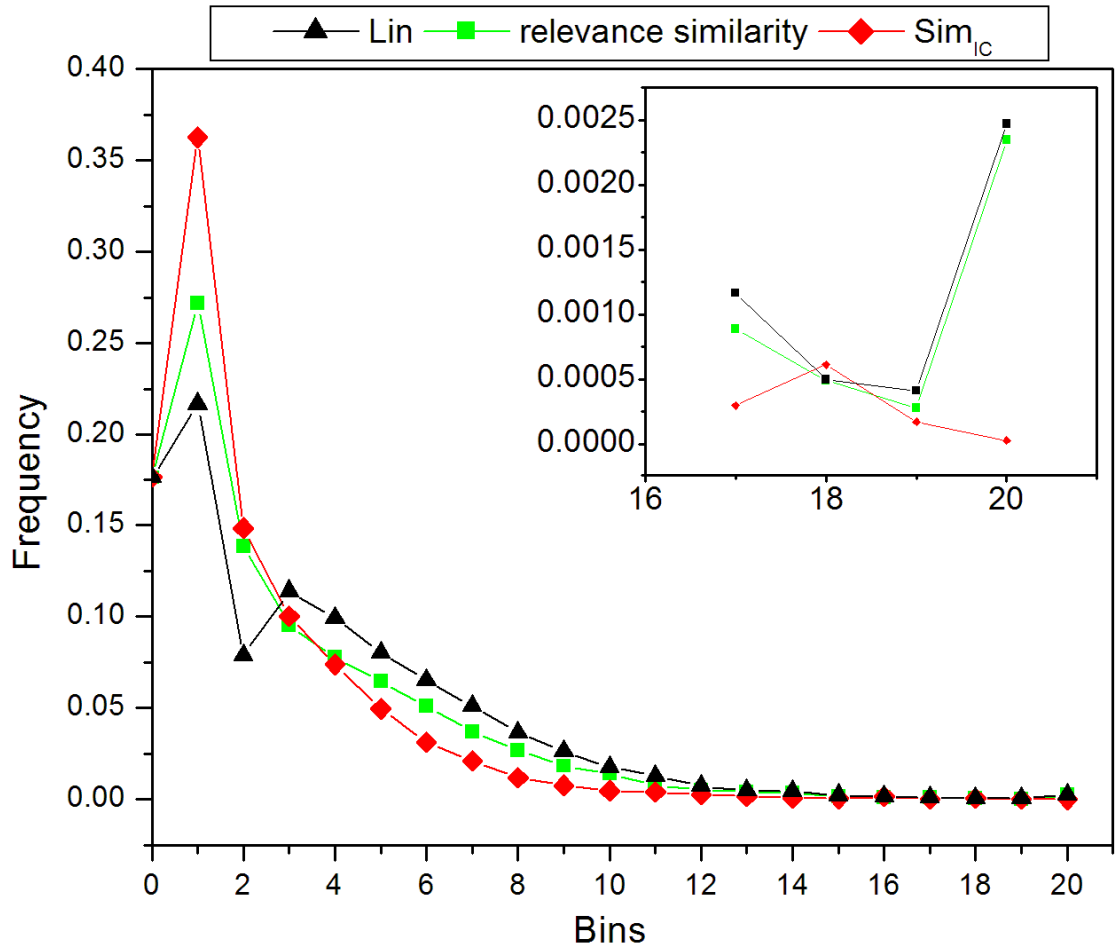

Figure 3

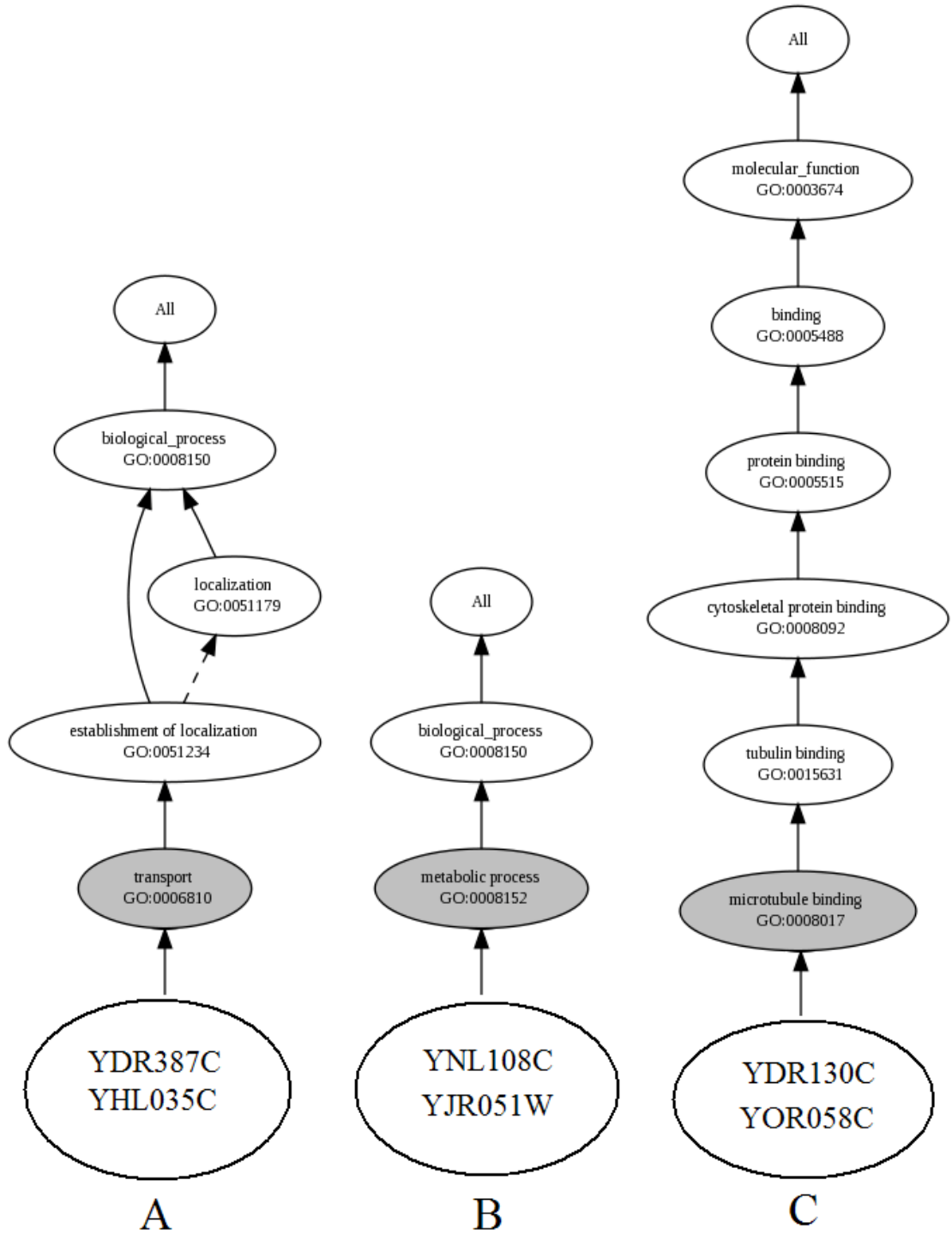

Figure 4

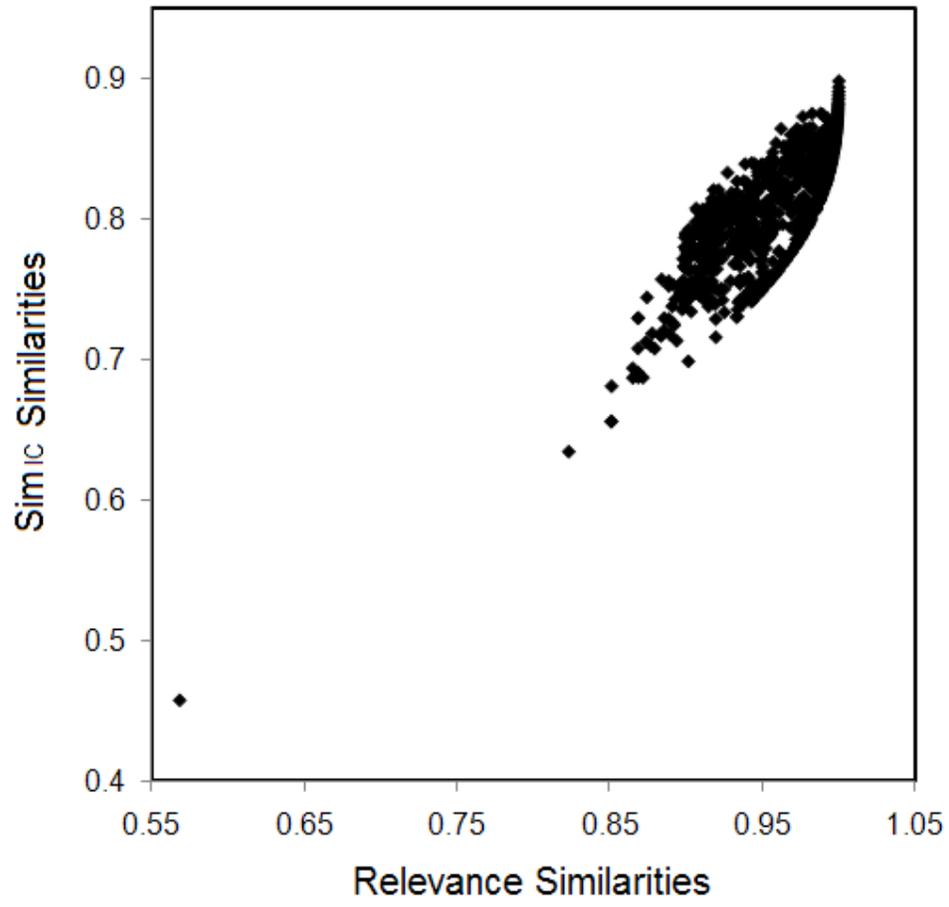

Figure 5

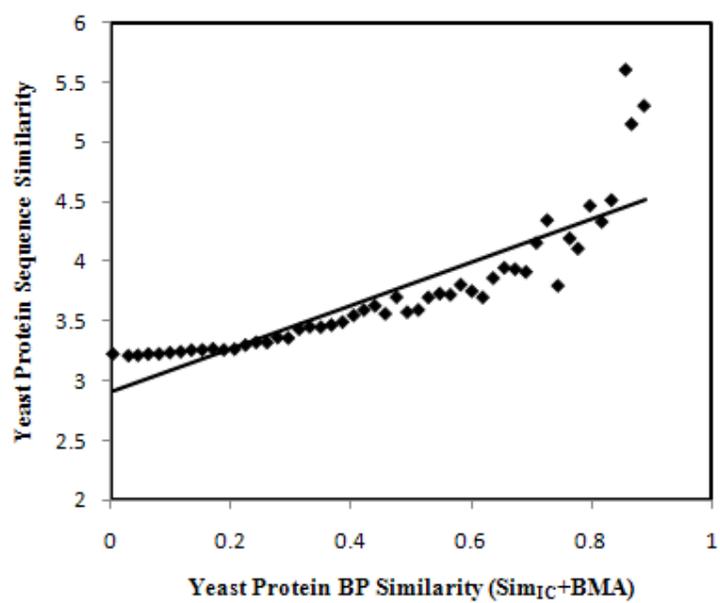

Figure 6

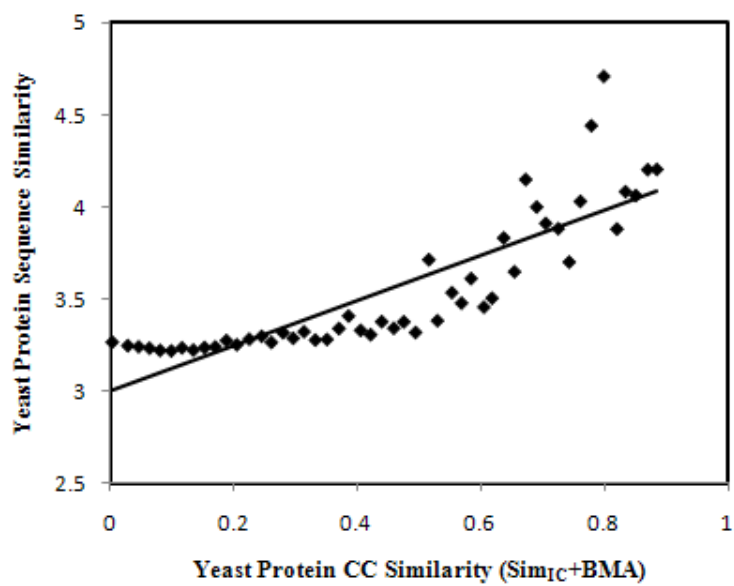

Figure 7

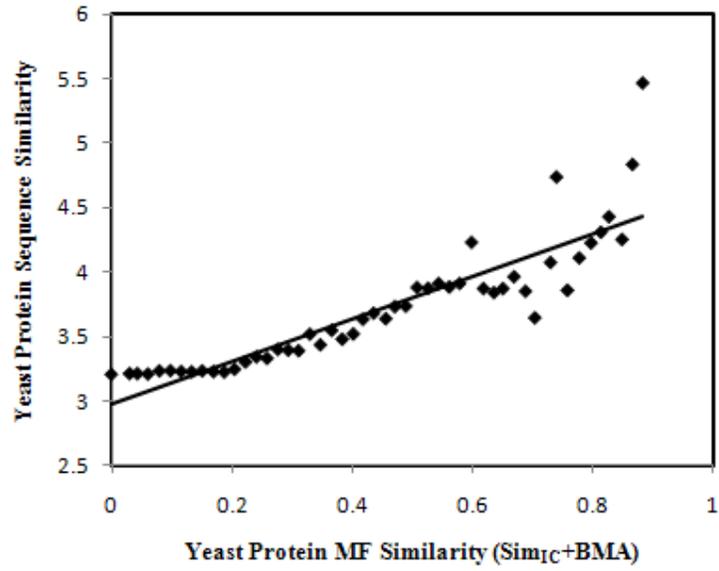

Figure 8

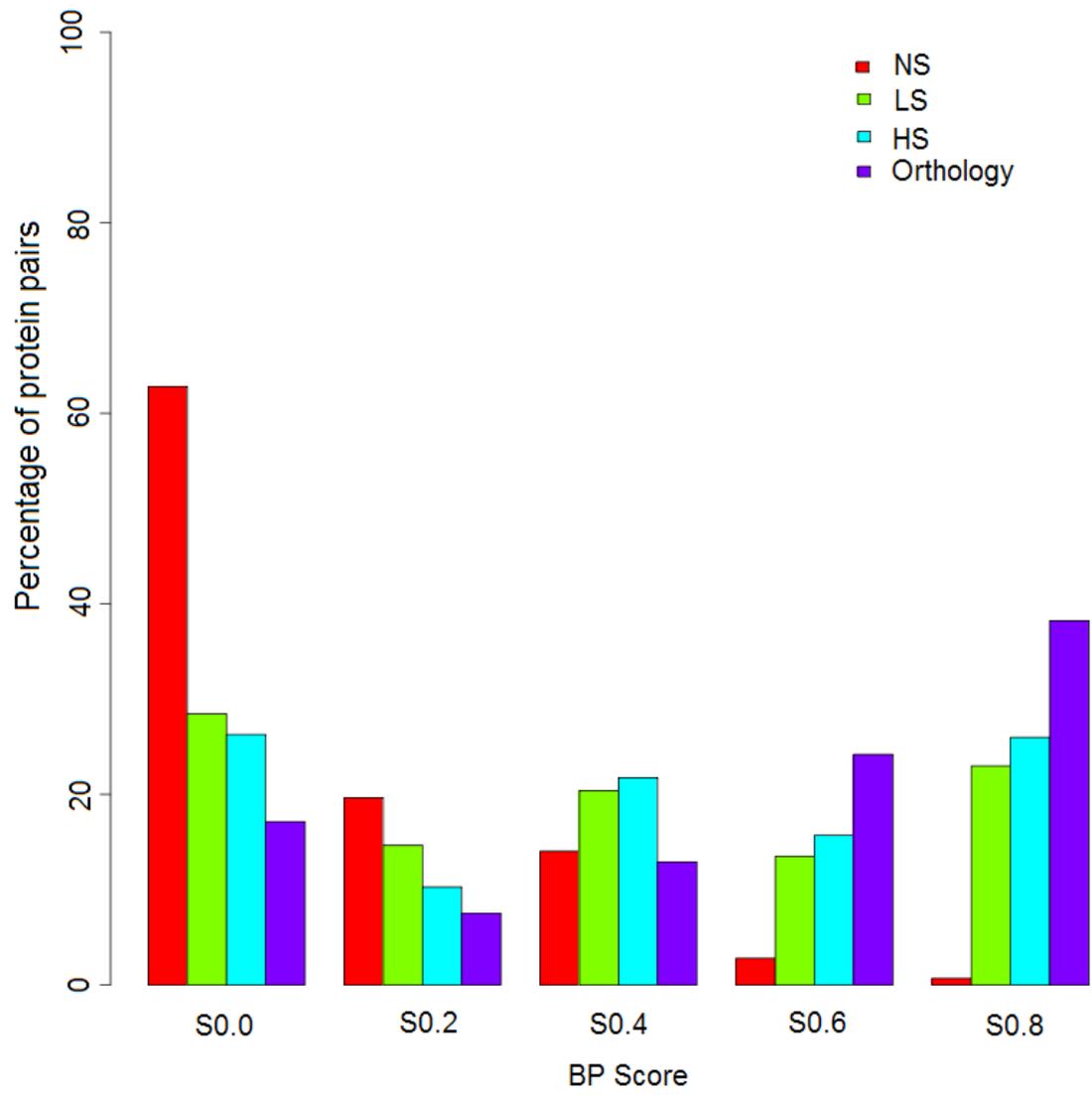

Figure 9

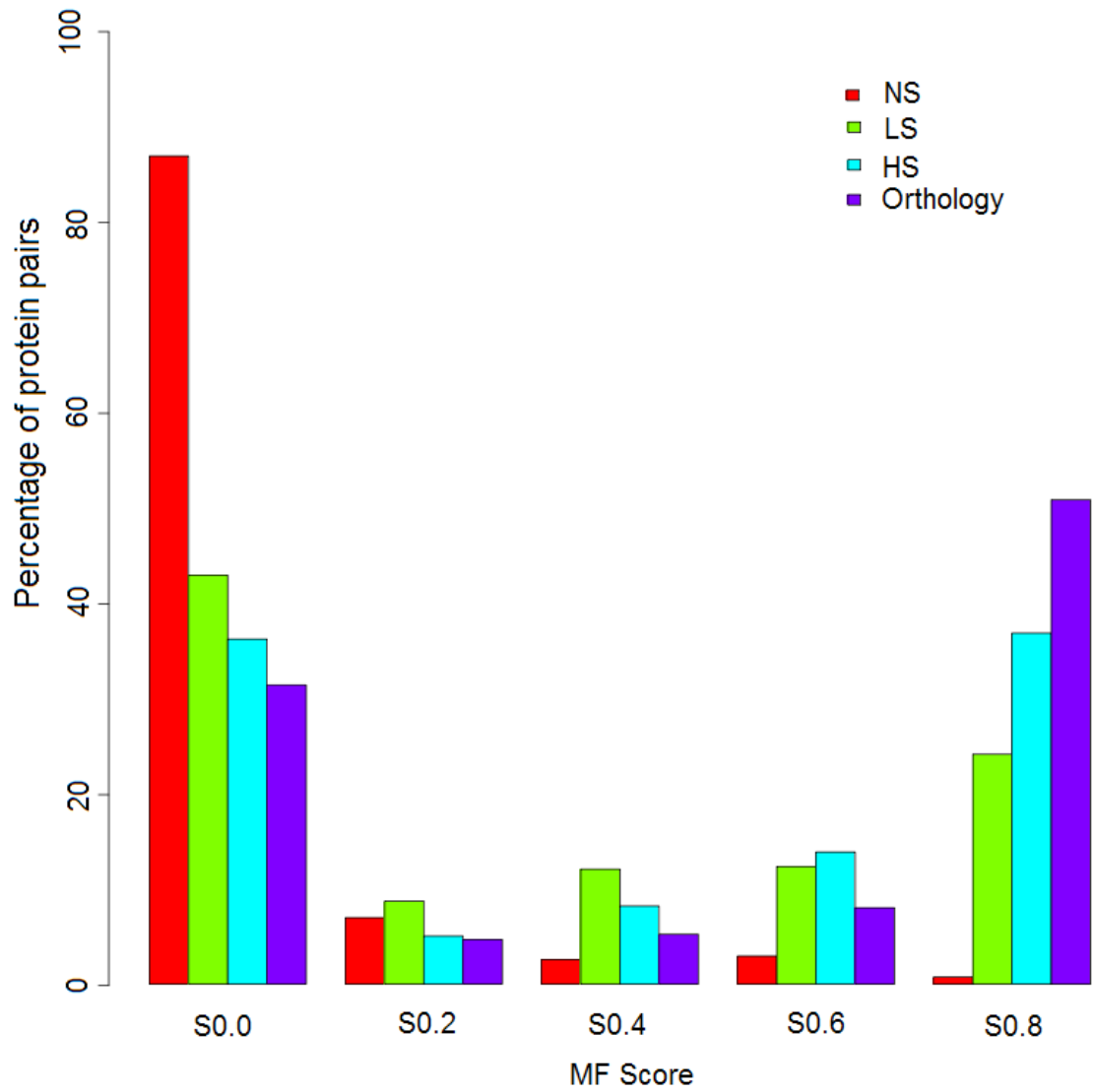

Figure 10

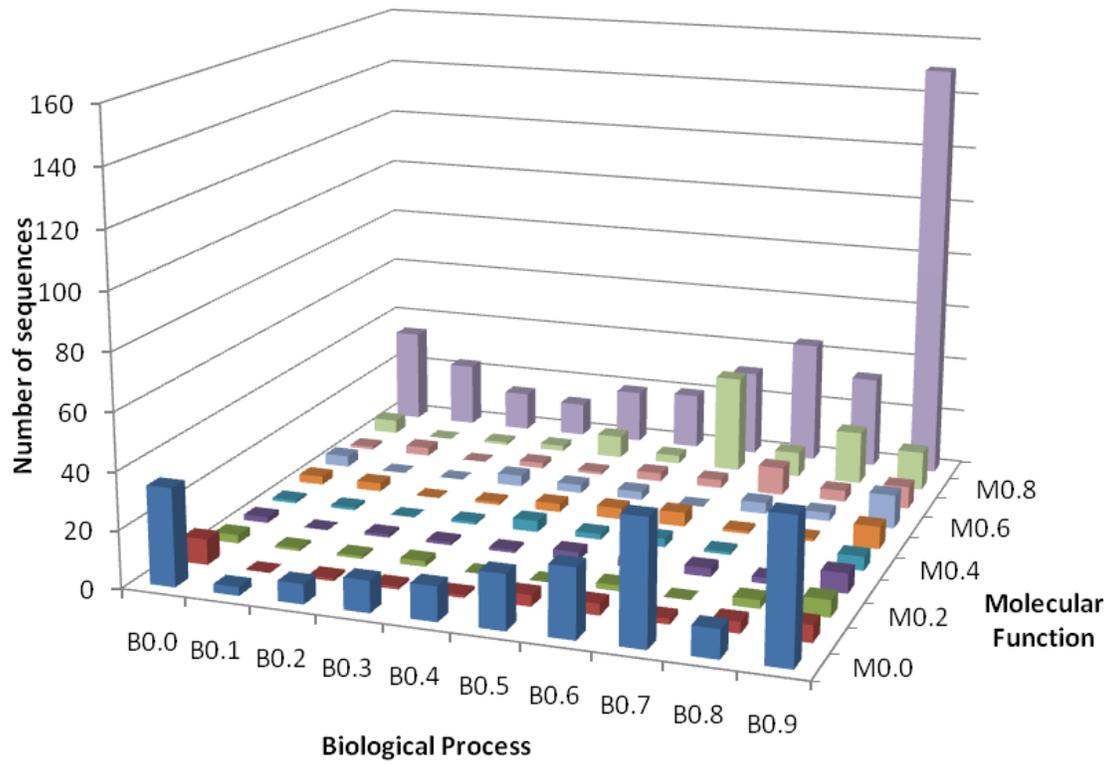

Figure 11

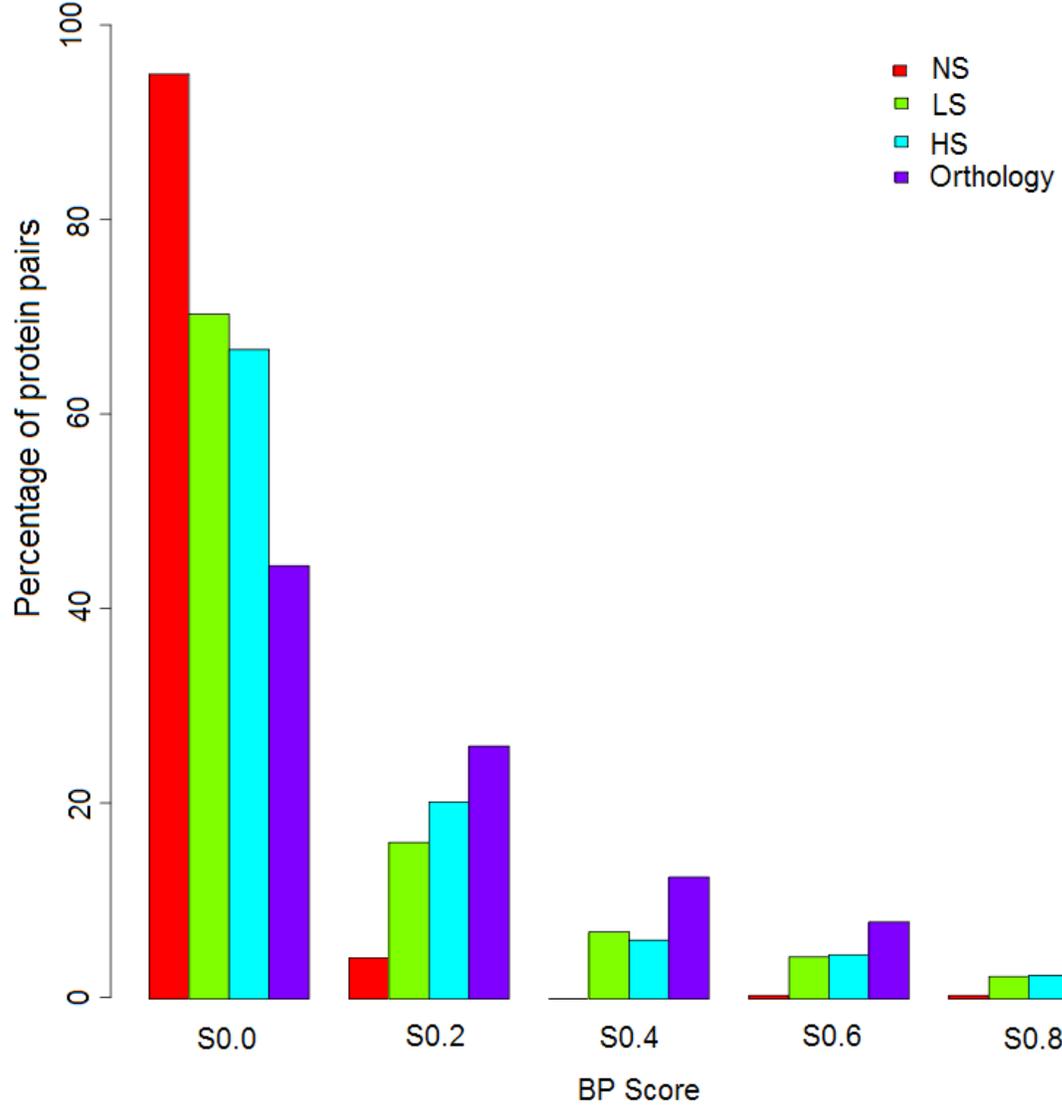

Figure 12

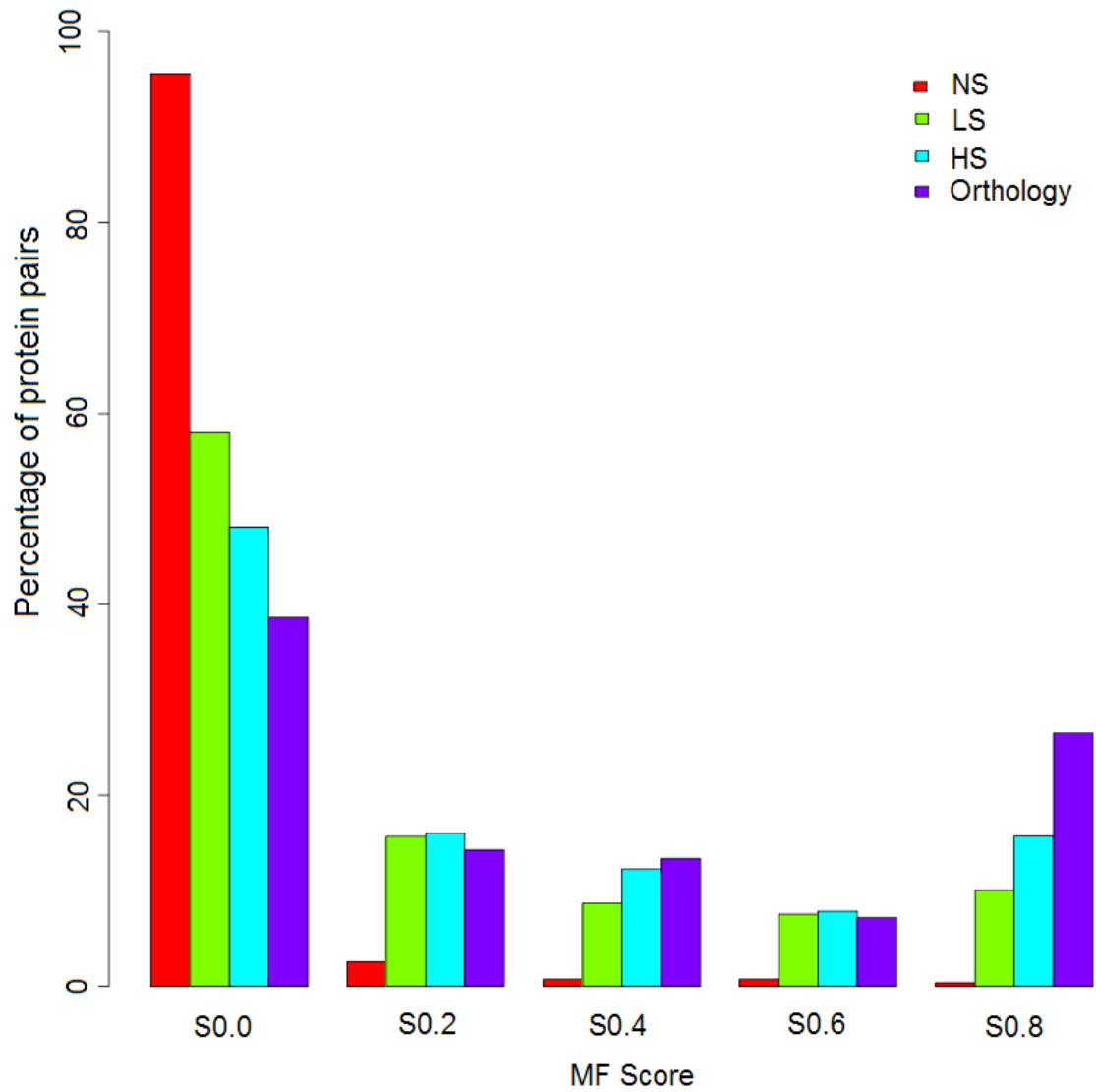

Figure 13

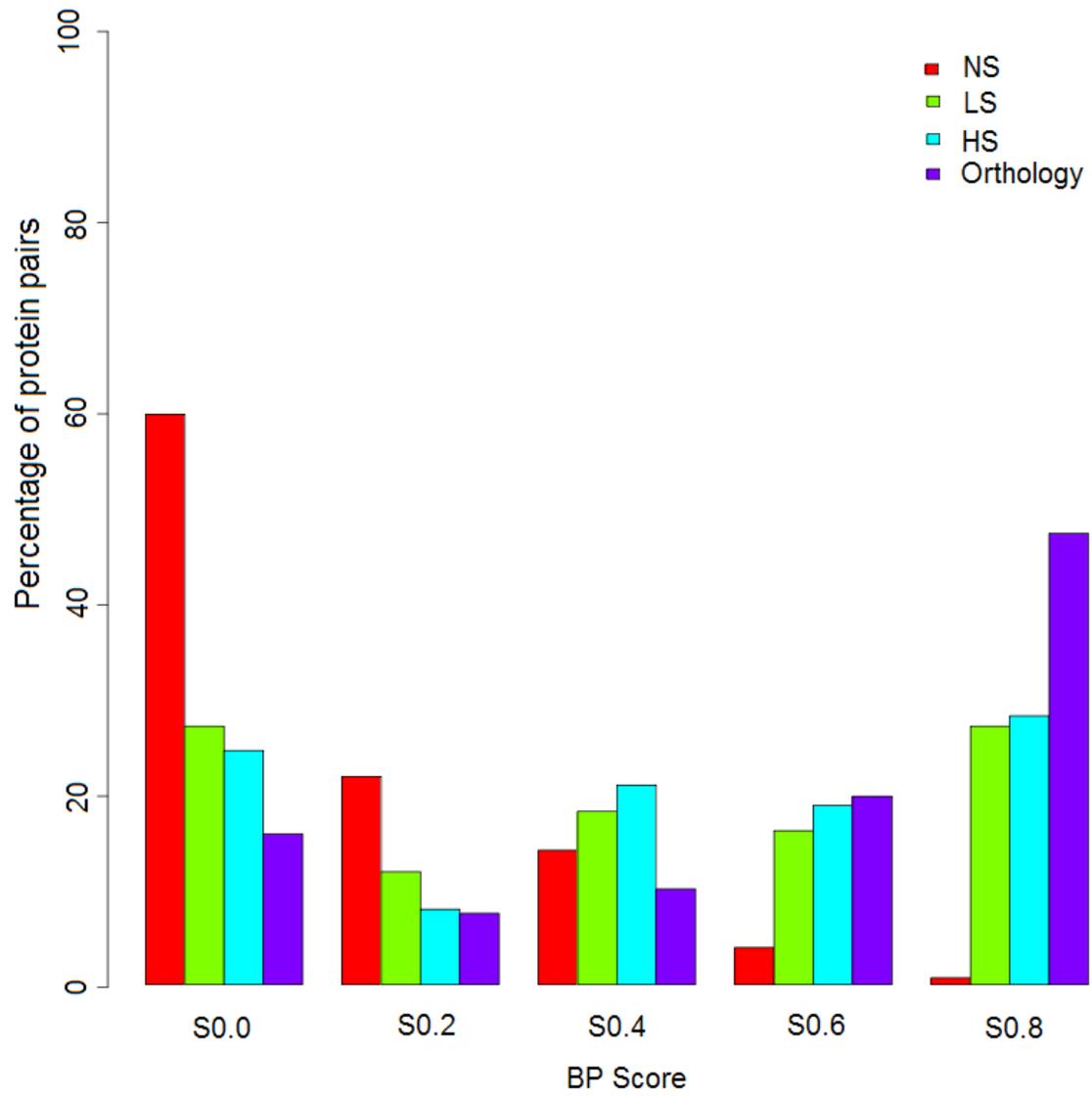

Figure 14

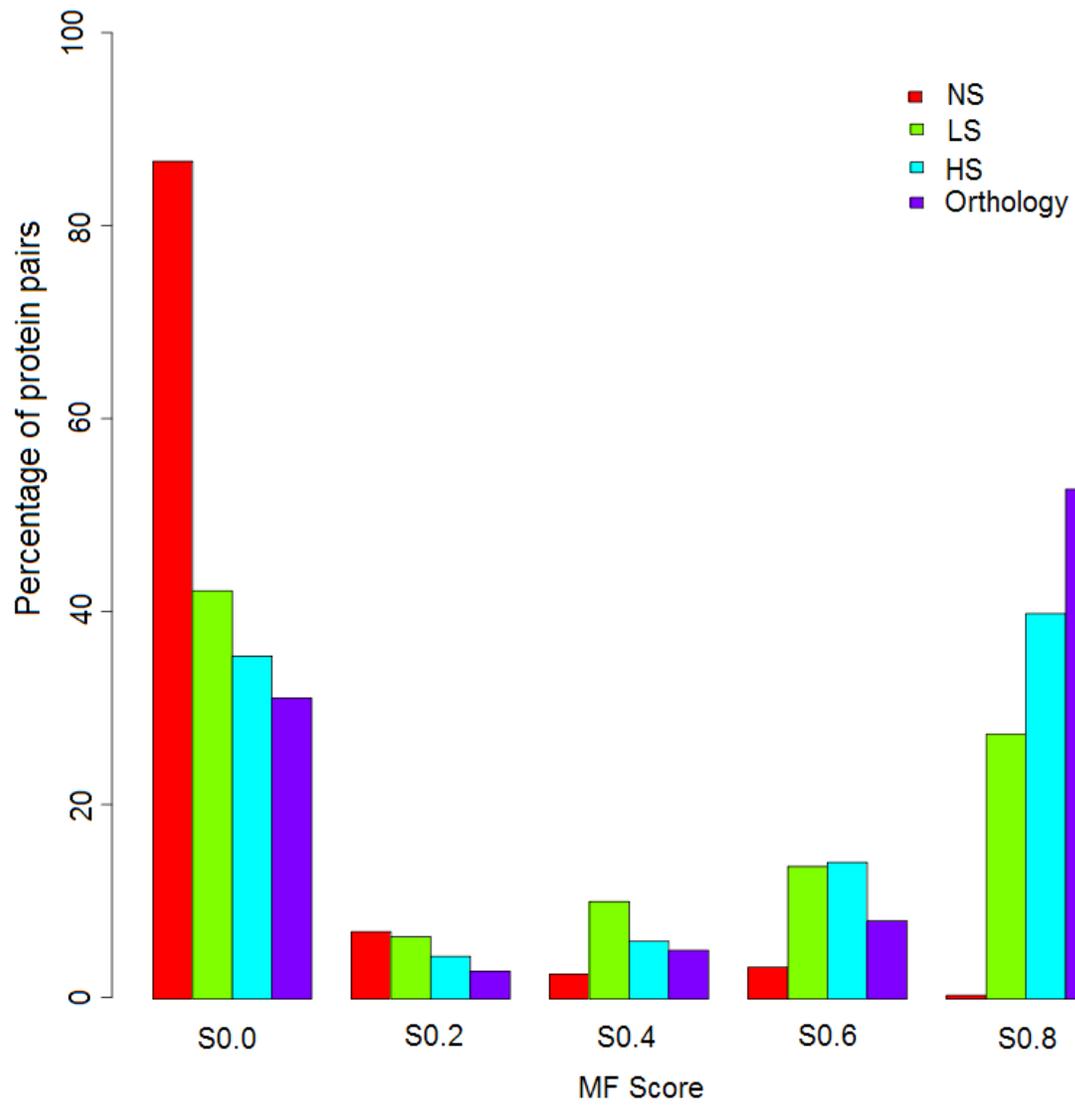

Figure 15

Figure 16

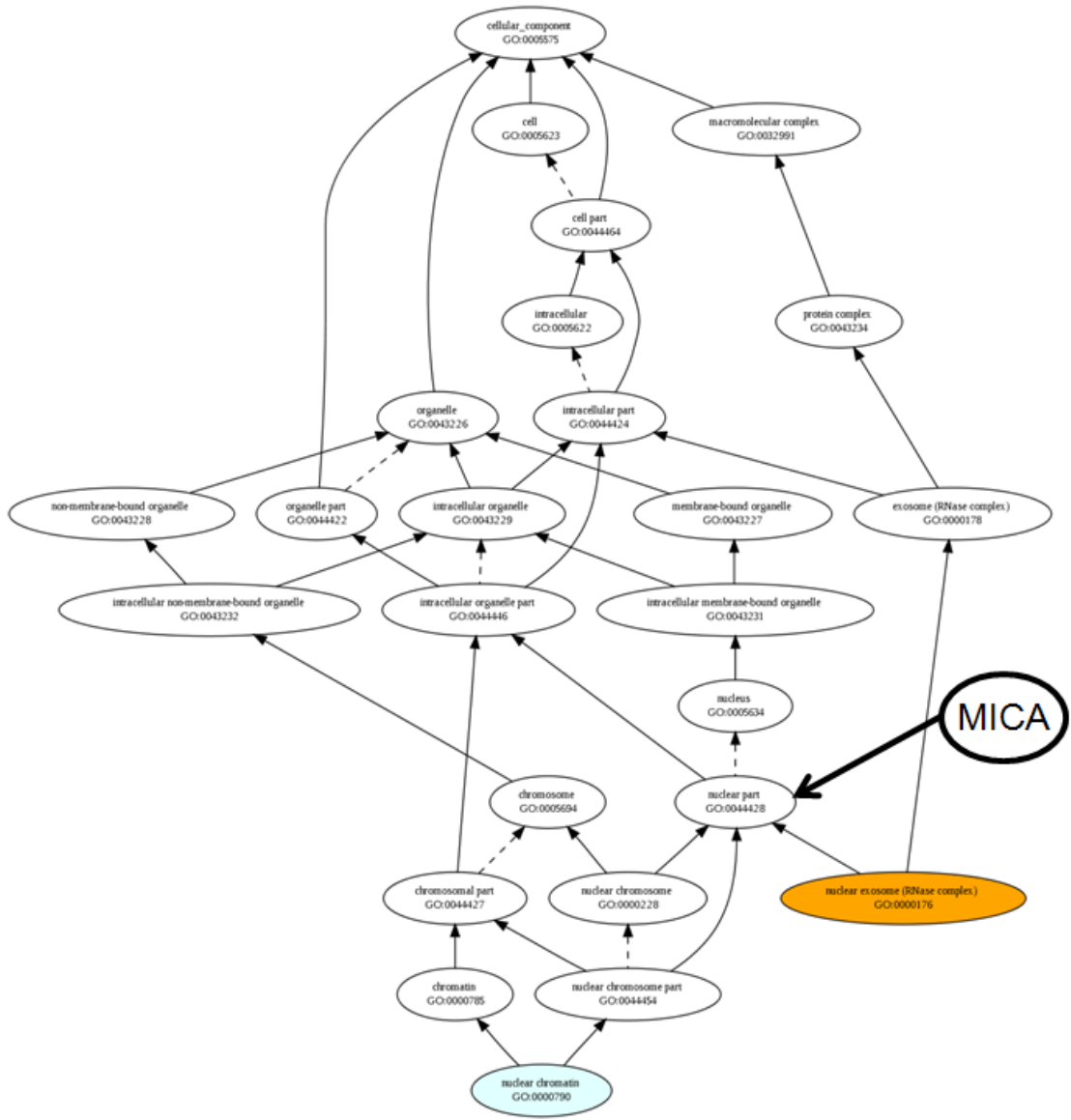

Figure 17

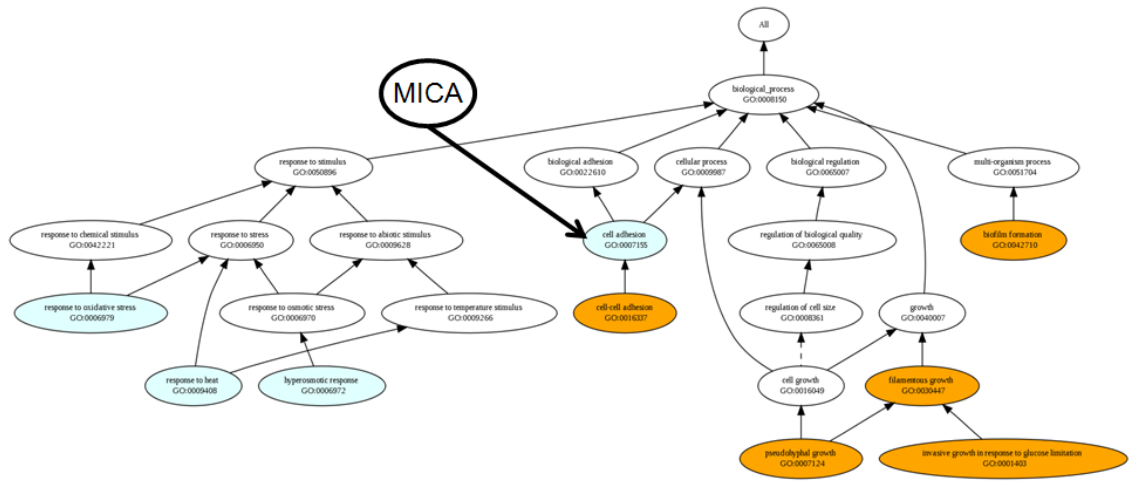

Figure 18

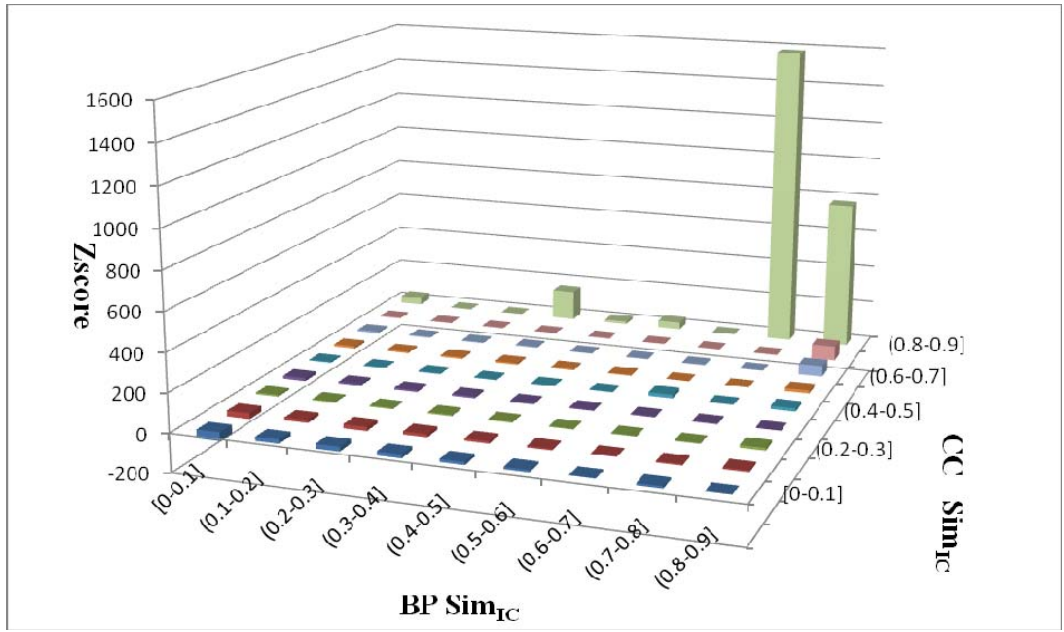

Figure 19

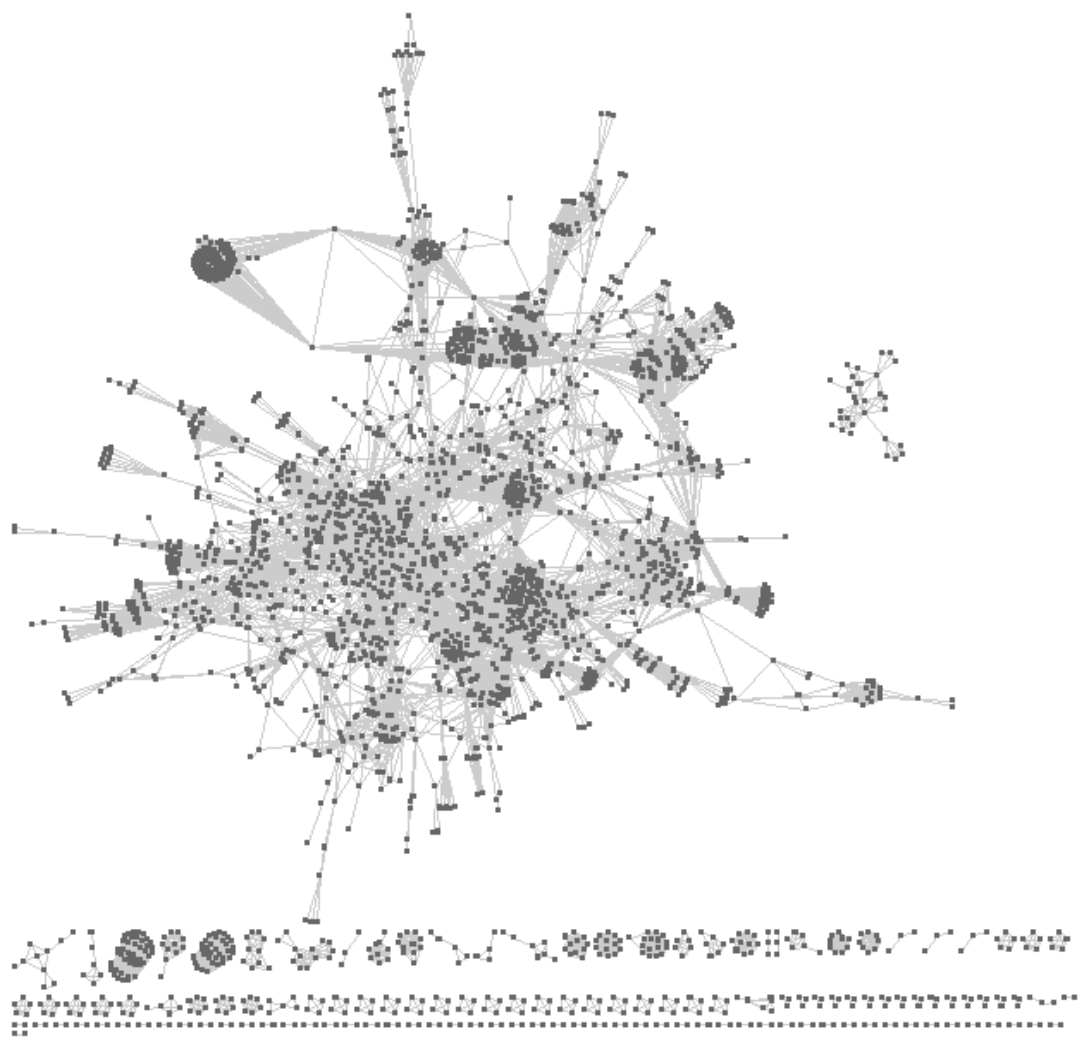

Figure 20

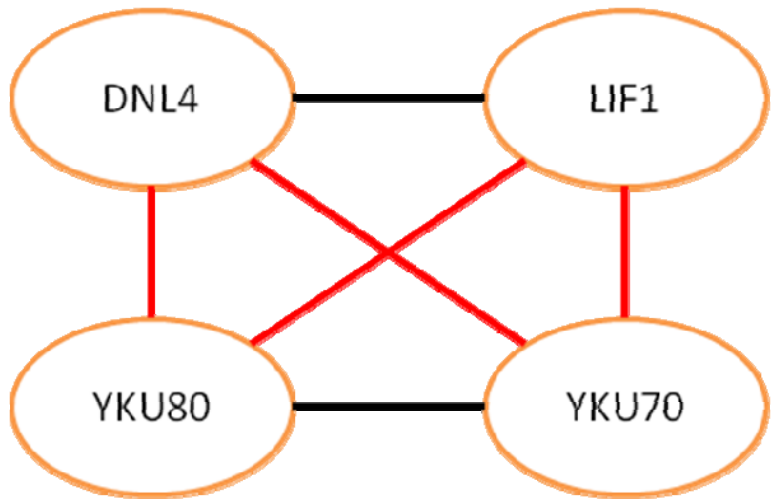